\documentclass[twocolumn,aps,prd,notitlepage,nofootinbib,
superscriptaddress,reprint,letterpaper]{revtex4}
\usepackage[usenames,dvipsnames]{xcolor} 
\usepackage{amsmath, latexsym}
\usepackage{float}
\usepackage{hyperref}
\usepackage[normalem]{ulem}
\usepackage{amssymb}
\usepackage{mathrsfs}
\usepackage{lmodern}
\usepackage[percent]{overpic}
\usepackage[T1]{fontenc}
\usepackage[utf8]{inputenc}
\usepackage{graphicx}
\usepackage{booktabs}
\usepackage{ragged2e}
\usepackage{tabularx}

\definecolor{orange}{rgb}{0.9,0.5,0}
\definecolor{maroon}{rgb}{.8,0,0}
\definecolor{purple}{rgb}{0.8,0.4,0.8}
\definecolor{gray}{rgb}{0.8242,0.8242,0.8242}
\definecolor{dodgerblue}{rgb}{0.12, 0.56, 1.0}
\definecolor{forestgreen}{rgb}{0.2, 0.6, 0.2}

\begin{document}

\title{The SpECTRE Cauchy-characteristic evolution system for rapid, precise waveform extraction}

\author{Jordan Moxon}
\affiliation{Theoretical Astrophysics, Walter Burke Institute for Theoretical Physics, California Institute of Technology, Pasadena, CA 91125, USA.}

\author{Mark A. Scheel}
\affiliation{Theoretical Astrophysics, Walter Burke Institute for Theoretical Physics, California Institute of Technology, Pasadena, CA 91125, USA.}

\author{Saul A. Teukolsky}
\affiliation{Theoretical Astrophysics, Walter Burke Institute for Theoretical Physics, California Institute of Technology, Pasadena, CA 91125, USA.}
\affiliation{Cornell Center for Astrophysics and Planetary Science,
  Cornell University, Ithaca, New York 14853, USA.}

\author{Nils Deppe}
\affiliation{Theoretical Astrophysics, Walter Burke Institute for Theoretical Physics, California Institute of Technology, Pasadena, CA 91125, USA.}

\author{Nils Fischer}
\affiliation{Max Planck Institute for Gravitational
    Physics (Albert Einstein Institute), Am M{\"u}hlenberg 1, D-14476 Potsdam,
    Germany}

\author{Francois H\'ebert}
\affiliation{Theoretical Astrophysics, Walter Burke Institute for Theoretical Physics, California Institute of Technology, Pasadena, CA 91125, USA.}

\author{Lawrence E. Kidder}
\affiliation{Cornell Center for Astrophysics and Planetary Science,
  Cornell University, Ithaca, New York 14853, USA.}

\author{William Throwe}
\affiliation{Cornell Center for Astrophysics and Planetary Science,
  Cornell University, Ithaca, New York 14853, USA.}


\newcommand{\todo}[1]{\textcolor{orange}{\texttt{TODO: #1}}} 
\providecommand{\JM}[1]{{\textcolor{forestgreen}{{#1}}}}
\providecommand{\MS}[1]{{\textcolor{Cerulean}{{#1}}}}
\providecommand{\ST}[1]{{\textcolor{dodgerblue}{{ST: #1}}}}
\newcommand{\Note}[1]{\textcolor{blue}{\textbf{[#1]}}}
\newcommand{\red}[1]{{\color{red}{#1}}}


\newcolumntype{Y}{>{\RaggedRight\arraybackslash}X}

\begin{abstract}
  We give full details regarding the new Cauchy-characteristic evolution (CCE) system
  in SpECTRE.
 The implementation is built to provide streamlined flexibility for either
  extracting waveforms during the process of a SpECTRE binary compact object
  simulation, or as a standalone module for extracting waveforms from worldtube
  data provided by another code base.
 Using our recently presented improved analytic formulation, the CCE system is
 free of pure-gauge logarithms that would spoil the spectral convergence of the
 scheme.  It gracefully extracts all five Weyl scalars, in addition to the news
and the strain.
 The SpECTRE CCE system makes significant improvements on previous implementations in
  modularity, ease of use, and speed of computation.
\end{abstract}
\maketitle

\section{Introduction}

Since the original gravitational wave detections by the LIGO-VIRGO
collaborations \cite{Abbott:2016blz, Abbott:2016nmj}, sensitivities of
ground-based detectors have continued to advance \cite{LIGOScientific:2018mvr,
Aasi:2013wya}.
A crucial requirement for the successful detection and parameter estimation of
astrophysical gravitational-wave sources is the accurate modelling of potential
gravitational wave signals.
Gravitational wave modelling is required both to construct templates for
extracting signals from instrumentation noise \cite{Hanna:2010zzb,
Flanagan:1997sx} and for performing follow-up parameter estimation
\cite{Abbott:2016apu,Kumar:2018hml,Kumar:2015tha, Lange:2017wki,
Lovelace:2016uwp}.  
Currently, the precision of numerical relativity waveforms is sufficient to
cause no significant bias in detections produced by the present generation of
gravitational wave detectors \cite{Purrer:2019jcp}.

As the technology of the current network of gravitational wave detectors
(Advanced LIGO \cite{InstrumentWhitePaper2020}, VIRGO, and KAGRA
\cite{Nguyen:2021zna}) continues to mature, next-generation ground based
interferometers (Cosmic Explorer \cite{reitze2019cosmic} and Einstein
Telescope \cite{Maggiore_2020}) are planned, and space-based gravitational wave
detector projects (LISA \cite{amaroseoane2017laser}, TianQin \cite{Luo_2016}
and DECIGO \cite{Sato_2017}) move forward, the demand for high-precision
waveform models for binary inspirals continues to grow.
Recent investigations \cite{Purrer:2019jcp} have indicated that future
ground-based gravitational wave detectors will have sufficient sensitivity that
current numerical relativity waveforms are not precise enough to produce
unbiased parameter recovery.   
Further, space-based gravitational wave detectors, such as LISA, will likely
observe several sources simultaneously, and sufficiently precise modelling of
each source will help make best use of the resulting data by improving the
capability to distinguish overlapping signals.

An important ingredient to improved precision for numerical relativity waveforms
is the refinement of waveform extraction methods. 
The process of waveform extraction refers to the calculation of the observable
asymptotic waveform from a strong-field simulation of the Einstein field
equations.  Current strong-field numerical relativity simulation methods are
`Cauchy' methods \cite{SpEC, Loffler:2011ay, Bruegmann:2003aw, Ruchlin:2017com}:
initial data is generated for a desired configuration of the compact binary
using an elliptic solve on a restricted region, and that spacelike hypersurface
data is evolved in the timelike direction.
 One output of a Cauchy simulation is the metric and its derivatives as a
function of time, evaluated on one or more spheres of finite distance from the
binary, typically $\sim 100 - 1000 M$ from the coalescence.
Waveform extraction then uses the Cauchy worldtube metric and its derivatives to
determine the observable asymptotic waveform that is directly applicable to
data analysis efforts for gravitational wave interferometers.

The most widely used technique of waveform extraction is the method of
extrapolation to large radii using several worldtubes of finite radius
\cite{Iozzo:2020jcu, Bishop:2016lgv}.
For each waveform quantity of interest, such as the gravitational wave strain or
one of the Weyl scalars, there is a clear power law asymptotic behavior in
well-behaved gauges.
The extrapolation method then fits for the leading behavior in $r^{-1}$ and
obtains a reasonable approximation for the asymptotic waveform.
The extrapolation method has been used to generate a great number of useful
waveforms for gravitational wave data analysis \cite{Boyle:2019kee,
Healy:2019jyf, Jani:2016wkt}.
However, the extrapolation method makes a number of simplifying assumptions
regarding the choice of coordinates and behavior of the field equations far from
the system that diminish the precision of the method.

In addition, there is good evidence \cite{Mitman:2020pbt} that there are large,
low-frequency parts of gravitational waveforms (`memory' contributions) that are
not well modeled
by waveform extrapolation. These memory effects do not have
significant impact on the frequency bands important for LIGO, but will likely be
important for more sensitive detectors (such as the Einstein Telescope or
Cosmic Explorer) or detectors sensitive to lower frequency bands (such as
DECIGO or LISA).

\begin{figure}
  \begin{overpic}[scale=0.6]{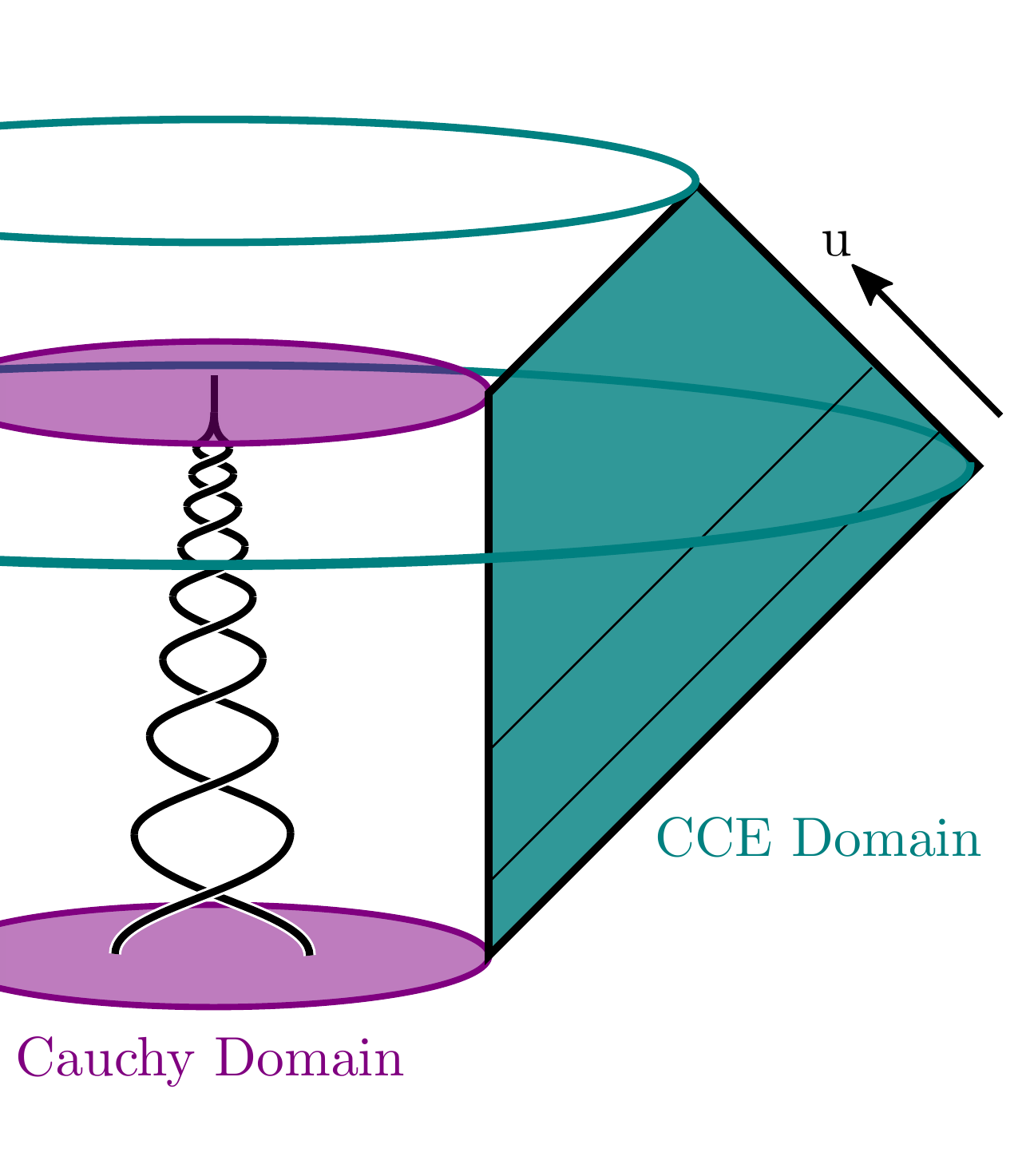}
    \put(61,84){\Large$\mathcal I^+$}
    \put(36,55){\Large$\Gamma$}
    \put(48,35){\Large$\Sigma_u$}
  \end{overpic}
  \caption{A sketch of the Cauchy and Characteristic domains. The Cauchy system evolves
    Einstein's equations on spacelike hypersurfaces,
    while the Characteristic system evolves Einstein's equations
    on compactified
    null hypersurfaces $\Sigma_u$ that extend
    to $\mathcal I^+$. Boundary conditions for the Characteristic system are required
      on the worldtube $\Gamma$ and are
      provided there by the Cauchy system.
  }
  \label{fig:cce_sketch}
\end{figure}

Cauchy-characteristic evolution\footnote{The acronym CCE has also been used in the past to refer to
  ``Cauchy-characteristic extraction'', which describes only the part of the
  computation moving from the Cauchy coordinates to a set of quantities that
  could separately be evolved on null characteristic curves.
  Most of our descriptions refer to the entire algorithm as a single part of the
  wave computation, so we refer to the combination of Cauchy-characteristic
  extraction and characteristic evolution as simply CCE.
} Cauchy-characteristic evolution (CCE)~\cite{Bishop:1996gt, Winicour:2012znc,
Babiuc:2011qi} is an alternative waveform extraction method that uses metric
data on a single worldtube $\Gamma$ to provide boundary conditions for a second full nonlinear
field simulation along hypersurfaces generated by outgoing null geodesics.
CCE avoids many of the assumptions made by other extraction methods, and instead
computes the full solution to Einstein's equations in a Bondi-Sachs coordinate
system at $\mathcal I^+$, from which waveform quantities may be unambiguously
derived.
The CCE domain and salient hypersurfaces are illustrated in Fig.~\ref{fig:cce_sketch}.

There are two notable previous implementations of CCE. The original
implementation, {\sc PITT Null} \cite{Winicour:1999sg, Bishop:1997ik}, is a part
of the Einstein Toolkit, and demonstrated the feasibility of the CCE approach.  Unfortunately,
as it is a finite difference implementation, {\sc PITT Null} struggles to achieve
high precision and can be very costly to run \cite{Barkett:2019uae}.
The first spectral implementation of CCE is a module of the Spectral Einstein
Code (SpEC). That implementation was first reported in \cite{Handmer:2014qha},
and has undergone a number of updates and refinements \cite{Handmer:2015dsa,
Handmer:2016mls}, including recent work that assembled a number of valuable
analytic tests that assisted in refining and optimizing the code
\cite{Barkett:2019uae}.

In this paper, we present our new implementation of CCE in the SpECTRE
\cite{deppe_nils_2021_5083825} code base, which incorporates a number of
improvements to the waveform extraction system.
The SpECTRE CCE module implements a modified version of the evolution system in
Bondi-Sachs coordinates \cite{Moxon:2020gha} that is able to guarantee that no
pure-gauge logarithms arise that spoil the spectral convergence of the scheme as
the system evolves. Further, the SpECTRE CCE
system is able to use formulation simplifications to implement the computation
for all five Weyl scalars as suggested in \cite{Moxon:2020gha}.
We have also implemented numerical optimizations specific to the SpECTRE CCE
system to ensure rapid and precise waveform extraction, and we have
re-implemented and extended the collection of tests that was previously
effective in testing and refining the SpEC implementation
\cite{Barkett:2019uae}.

SpECTRE \cite{Kidder:2016hev, deppe_nils_2021_5083825} is a next-generation code base for which the aim
is to construct scalable multi-physics simulations of astrophysical phenomenon
such as neutron star mergers, binary black hole coalescences, and core-collapse
supernovae.
It is the goal of the SpECTRE project to construct a highly precise
astrophysical simulation framework that scales well to
${}\gtrsim 10^6$ cores.
The core SpECTRE evolution system uses discontinuous Galerkin methods with a
task-based parallelism model.
The discontinuous Galerkin method has the ability to refine a domain by
subdividing the computation into local calculations coupled by boundary fluxes.
SpECTRE then uses the task-based parallelism framework, {\sc charm++}
\cite{CharmppOOPSLA93,CharmAppsBook:2013,Charm}, to schedule and run the
resulting multitude of separate calculations, which ensures good scaling
properties of the method.

The CCE system in SpECTRE enjoys some efficiency gain from sharing a common
well-optimized infrastructure with the discontinuous Galerkin methods and makes
modest use of the parallelization framework (see Sec.~\ref{sec:parallelization_modularity}).
However, the characteristic evolution itself is implemented as a single spectral
domain that covers the entire asymptotic region from the worldtube $\Gamma$ out
to $\mathcal I^+$. The smooth behavior of the
metric away from the binary coalescence ensures exponential convergence of the
monolithic spectral method. 
In principle, the CCE method could be applied to a subdivided asymptotic domain.
However, the unusual features of the field equations for CCE (reviewed in
Sec.~\ref{sec:evolution_system}) would require special treatment to
appropriately account for boundary information.
Moreover, any subdivision of the angular direction would obscure the spherical
shell geometry that permits efficient calculation of the angular degrees of
freedom of the system via spin-weighted spherical harmonic (SWSH) methods.
 
It is important to note that the SpECTRE CCE module, like every part of SpECTRE,
is a rapidly evolving open-source code base.
The discussion in this paper represents as completely as possible the state of
our efforts to optimize and refine the system at the time of publication.
However, we will continue to make modifications and improvements, so we
encourage the reader to explore the full code base at \cite{SpECTRE_github}, and
refer to the documentation at \cite{SpECTRE_dox}.
For up-to-date details on making use of the standalone SpECTRE CCE system,
please see the documentation page \cite{SpECTRE_dox_cce_tutorial}.

We first describe the mathematical aspects of the evolution system, including
the incorporation of formulation improvements from \cite{Moxon:2020gha} in
Sec.~\ref{sec:evolution_system}. Next, we discuss some of the numerical methods
that we have constructed for our new SpECTRE implementation to improve runtime
and precision in Sec.~\ref{sec:numerical_optimizations}. We discuss the how the
SpECTRE CCE module fits into the wider task-based SpECTRE infrastructure in
Sec.~\ref{sec:parallelization_modularity}. Finally, we demonstrate the precision
and accuracy of the code by applying the system to a collection of analytic test
cases in Sec.~\ref{sec:battery_of_tests}, and to a realistic use-case of
extracting data from a binary black-hole evolution from SpEC in
Sec.~\ref{sec:bbh_trials}. We describe the major future improvements that we
hope to make for the CCE system in Sec.~\ref{sec:upcoming_improvements}.

\section{The evolution system} \label{sec:evolution_system}

The discussion of CCE and its numerical implementations relies closely on a
number of coordinate systems.  We use the following notation for coordinate
variables and spacetime indices:
\begin{itemize}
\item $x^{\alpha}$: $\{u, r, \theta, \phi\}$
  are generic Bondi-like coordinates.  These are the coordinates determined by
  the first stage of local coordinate transformations at the worldtube first
  derived in \cite{Bishop:1998uk}.
\item $\hat x^{\hat \alpha}$: $\{\hat u, \hat r, \hat \theta, \hat \phi\}$
  are partially flat Bondi-like coordinates introduced in \cite{Moxon:2020gha}.
\item $\breve x^{\breve \alpha}$: $\{\breve u, \breve y, \breve \theta, \breve \phi\}$
  are numeric partially flat coordinates. These are the coordinates directly
  represented in the SpECTRE numeric implementation, and are related to the
  partially flat Bondi-like coordinates by
  \begin{subequations}
  \begin{align}
    \breve u &= \hat u,\hspace{1cm}
    \breve y = 1 - 2\hat R/\hat r,\hspace{1cm}\notag\\
    \breve \theta &= \hat \theta,\hspace{1cm}
    \breve \phi = \hat \phi,
  \end{align}
  \end{subequations}
  where the worldtube hypersurface is determined
  by $\hat r = \hat R(\hat u, \hat \theta, \hat \phi)$.

\item $\mathring x^{\mathring \alpha}$ : $\{\mathring u, \mathring r, \mathring
    \theta, \mathring \phi\}$ are the asymptotically flat `true' Bondi-Sachs
    coordinates. These are the coordinates in which we'd like to determine the final
    waveform quantities.
\end{itemize}
We use Greek letters $\alpha, \beta, \gamma, \dots$ to represent spacetime
indices, uppercase roman letters $A, B, C, \dots$ to represent spherical
angular indices,
and lowercase roman letters from the middle of the alphabet
$i,j,k,\dots$ to represent spatial indices.

When relevant, we similarly adorn the spin-weighted scalars and tensors that
represent components of the metric to indicate the coordinates in which they are
components of the Bondi-like metric. For instance, the $g_{\hat r \hat u}$
component of a partially flat Bondi-like metric is $-e^{2 \hat \beta}$.
Our notation conventions are consistent with our previous paper regarding the
mathematics of the CCE system \cite{Moxon:2020gha}.

\subsection{Spectral representation}

The SpECTRE CCE system represents its null hypersurface data on the domain
$I\times S^2$, where the real interval $I$ describes the domain $y \in [-1, 1]$
for compactified radial coordinate
\begin{equation}
  \breve y = 1 - \frac{2 \hat R(\hat u, \hat x^{\hat A})}{\hat r},
\end{equation}
where $\hat r$ is the partially flat Bondi-like radial coordinate and $\hat R$
is the Bondi-like radius at the worldtube.

We use a pseudospectral representation for each physical variable on this domain,
using Gauss-Lobatto points for the radial dependence, and
{\sc libsharp}\cite{Reinecke:2013, Libsharp}-compatible 
collocation points for the angular dependence.
The angular collocation points are chosen to be equiangular in the $\phi$
direction, and Gauss-Legendre points in $\cos\theta$~\footnote{It
  is of some numerical convenience that there are no points at the poles,
  where spherical polar coordinates are singular. However, care must still be
  taken to avoid unnecessary factors of $\sin\theta$ in quantities like
 derivative operators, as they give rise to
  greater numerical errors when points are merely \emph{close} to the pole.}.

The choice of Gauss-Lobatto points for the radial dependence
 simplifies the CCE algorithm
because it is convenient to specify boundary conditions for the
radial integrals as simple boundary values.

The choice of angular collocation points enables fast SWSH transforms, so that
{\sc libsharp} routines can efficiently provide the angular harmonic
coefficients ${}_s a_{lm}(\breve y)$ for an arbitrary function $f(\breve y, \breve\theta, \breve\phi)$ of spin weight $s$, defined by
\begin{equation}
  f(\breve y, \breve\theta, \breve\phi) = \sum_{\ell m} {}_s a_{\ell m}(\breve y) {}_s Y_{\ell m}(\breve\theta, \breve\phi).
\end{equation}
Here ${}_s Y_{\ell m}(\breve\theta, \breve\phi)$ are the SWSHs as defined in Eq.~(\ref{eq:SwshToWigner}).

We then perform all angular calculus operations using the spin-weighted
derivative operators $\breve \eth$ and $\breve{\bar \eth}$. We use an angular
dyad $\breve q^{\breve A}$:
\begin{equation}
  \breve q^{\breve A} = \left\{-1, \frac{-i}{\sin \breve \theta}\right\}.
\end{equation}
Then, for any spin-weighted scalar quantity $\breve v = \breve q_1^{\breve A_1}
\dots \breve q_n^{\breve A_n} \breve v_{\breve A_1 \dots \breve A_n}$, where
each $\breve q_i$ may be either $\breve q$ or $\breve{\bar q}$, we define the
spin-weighted derivative operators
\begin{subequations}
\begin{align}
  \breve \eth \breve v = \breve q_1^{\breve A_1} \dots \breve q_n^{\breve A_n} \breve q^{\breve B} \breve D_{\breve B} v_{\breve A_1 \dots \breve A_n}, \\
  \breve{ \bar \eth} \breve v = \breve q_1^{\breve A_1} \dots \breve q_n^{\breve A_n} \breve{\bar q}^{\breve B} \breve D_{\breve B} \breve v_{\breve A_1 \dots \breve A_n},
\end{align}
\end{subequations}
where $\breve D_{\breve A}$ is the angular covariant derivative.
All angular derivatives may be expressed in a combination of $\breve \eth$ and
$\breve {\bar \eth}$ operators.
We perform angular differentiation of an arbitrary
  function $f(\breve y, \breve\theta, \breve\phi)$ of spin weight $s$
by transforming to SWSH modes on each concentric
spherical slice of the domain represented by ${}_s a_{l m}(\breve y)$, then
applying the diagonal modal multipliers
\begin{subequations} \label{eq:spin_weighted_derivative_factors}
\begin{align}
  \breve  \eth& f(\breve y, \breve \theta, \breve \phi) \notag\\
      & = \sum_{\ell m} \sqrt{(\ell - s) (\ell + s + 1)} {}_s a_{\ell m}(\breve y)\; {}_{s + 1} Y_{\ell m}(\breve \theta, \breve \phi)\\
  \breve{ \bar \eth}& f(\breve y,\breve \theta,\breve \phi) \notag\\
      &= \sum_{\ell m} -\sqrt{(\ell + s)(\ell - s + 1)} {}_s a_{\ell m}(\breve y)\; {}_{s - 1} Y_{\ell m}(\breve\theta, \breve\phi),
\end{align}
\end{subequations}
and then performing an inverse transform.

In addition, it is occasionally valuable to apply the inverse of the angular
derivative operators $\breve\eth$ and $\breve{\bar \eth}$.
This can be performed applying the inverse of the multiplicative factors in the
modal representation (\ref{eq:spin_weighted_derivative_factors}), and is
approximately as efficient to compute as the derivative.

\subsection{Hierarchical evolution system}

For evolution in the characteristic domain (see Fig.~\ref{fig:cce_sketch}), we solve the Einstein field equations for
the spin-weighted scalars that appear in the Bondi-Sachs form of the metric:
\begin{align} \label{eq:bs_metric}
  ds^2 =& -\left(e^{2  \beta} \frac{ V}{ r} -  r^2  h_{ A  B} U^{ A}  U^{ B}\right) d u^2 - 2 e^{2  \beta} d u d r \notag\\
  &- 2  r^2  h_{ A  B}  U^{ B} d u d x^{ A} +  r^2  h_{ A  B} d x^{ A} d x^{ B}.
\end{align}
The spin-weighted scalars that are used in the evolution system are then
$J, \beta, Q, U, W,$ and $H$, where
\begin{subequations}
\begin{align}
  U &\equiv U^A q_A,\\
  Q &\equiv r^2 e^{-2 \beta} q^A h_{A B} \partial_r U^B,\\
  r^2 W &\equiv V - r,\\
  J &\equiv \frac{1}{2} q^A q^B h_{A B},\\
  K &\equiv \frac{1}{2} q^A \bar q^B h_{A B}.
\end{align}
\end{subequations}
In a Bondi-like metric, surfaces of constant $u$ are generated by outgoing
null geodesics.  The Bondi-Sachs metric further imposes asymptotic conditions on
each component of the metric that we will not impose for all of our coordinate
systems. The same form \eqref{eq:bs_metric} holds in any Bondi-like coordinates,
including the partially flat Bondi-like coordinates $\hat x^{\hat \alpha}$ and
true Bondi-Sachs coordinates $\mathring x^{\mathring \alpha}$.

It is important to note that for numerical implementations, the system is
usually not evolved in a true Bondi-Sachs coordinate system.
For convenience of numerical calculation, most CCE implementations enforce gauge
choices only at the worldtube boundary, and therefore do not ensure asymptotic
flatness.
The SpECTRE CCE implementation employs
a somewhat different tactic, as the generic
Bondi-like gauge is vulnerable to pure-gauge logarithmic dependence that
spoils spectral convergence.
Instead, we use the partially flat gauge introduced in \cite{Moxon:2020gha},
which ensures that the evolved coordinates are in the asymptotically inertial
angular coordinates, while keeping the time coordinate choice fixed by the
arbitrary Cauchy time coordinate.

In the Bondi-like coordinates, it is possible to choose a subset of the Einstein
field equations that entirely determine the scalars $\{J,\beta,U,W\}$ and
that form a computationally elegant, hierarchical set of differential
equations.  Represented
in terms of the numerical Bondi-like coordinates $\{\breve u, \breve y, \breve
\theta, \breve \phi\},$ the hierarchical differential equations take the form
\begin{subequations} \label{eq:hierarchy}
\begin{align}
  \partial_{\breve y} \breve \beta &= S_{\breve \beta}(\breve J),\\
  \partial_{\breve y}((1-\breve y)^2 \breve Q) &= S_{ \breve Q}(\breve J,\breve  \beta),\\
  \partial_{\breve y} \breve  U &= S_{\breve U}(\breve J,\breve  \beta,\breve  Q),\\
  \partial_{\breve y}((1-\breve y)^2 \breve W) &= S_{\breve W}(\breve J, \breve \beta, \breve Q, \breve U),\\
  \big[\partial_{\breve y}((1-\breve y) \breve H) +& L_{\breve H}(\breve J, \breve \beta, \breve Q, \breve U, \breve W) \breve H\notag\\
  +& L_{\breve {\bar H}}(\breve J, \breve \beta, \breve Q, \breve U, \breve W) \breve{\bar H}\big]\notag\\
                                   &=S_{\breve H}(\breve J,\breve \beta,\breve Q,\breve U,\breve W),\\
  \partial_{\breve u} \breve J &= \breve H. \label{eq:hierarchy_du_j}
\end{align}
\end{subequations} The detailed definitions for the source functions $\breve
S(\dots)$ and the factors $L_{\breve H}$ in \eqref{eq:hierarchy} can be found in
Sec.~IV of \cite{Moxon:2020gha}.
We emphasize that the only time derivative appearing in the core evolution
system \eqref{eq:hierarchy} is that of $\breve J$
\eqref{eq:hierarchy_du_j}, so we have only the single complex field to evolve
and all of the other equations are radial constraints within each null
hypersurface.

The SpECTRE CCE system requires input data specified on two hypersurfaces: the
worldtube $\Gamma$ and the initial hypersurface $\Sigma_{\breve u_0}$
(see Fig.~\ref{fig:cce_sketch}).
The worldtube surface data must provide sufficient information to set the
boundary values for each of the radial differential equations in \eqref{eq:hierarchy}.
Namely, we must specify $\breve \beta$, $\breve U$, $\breve Q$, $\breve W$, and
$\breve H$ at the worldtube (see Sec. \ref{sec:gauge_corrected} below).
The worldtube data is typically specified by determining the full spacetime
metric on a surface of constant coordinate radius in a Cauchy code, then
performing multiple gauge transformations to adapt the boundary data to the
appropriate partially flat Bondi-like gauge.

The initial hypersurface data requires specification only of the single evolved
field $\breve J$.
In contrast to Cauchy approaches to the Einstein field equations, the initial
data for CCE does not have a collection of constraints that form an elliptic
differential equation.
Instead, $\breve J$ may be arbitrarily specified on the initial data surface,
constrained only by asymptotic flatness conditions.
The choice of ``correct'' initial data to best match the physical history of an
inspiral system, however, remains very difficult.
We discuss our current heuristic methods for fixing the initial hypersurface
data in Sec.~\ref{sec:initial_data}.

\subsection{Gauge-corrected control flow} \label{sec:gauge_corrected}

The SpECTRE CCE system implements the partially flat gauge strategy
discussed at length in \cite{Moxon:2020gha}.
The practical impact of the method is that we must include the evolved angular
coordinates in the process of determining the Bondi-Sachs scalars for the radial
hypersurface equations.
Past implementations have performed the angular transformation at
$\mathcal I^+$, which results in a simpler algorithm, but also gives rise to
undesirable pure-gauge logarithmic dependence.

In this discussion, we make use of the local Bondi-Sachs-like coordinates
$\hat x^{\hat \mu}$ on the worldtube that are determined by the standard
procedure introduced in \cite{Bishop:1996gt} and reviewed in
\cite{Barkett:2019uae,Moxon:2020gha}. This procedure
obtains a unique Bondi-Sachs-like
coordinate system by generating a null hypersurface with geodesics outgoing with
respect to the worldtube, and with time and angular coordinates chosen to match
the Cauchy coordinates on the worldtube.

In the below discussion, we make use of an intermediate spin-weight 1 scalar
\begin{equation}
  \mathcal U = \breve U + \mathcal U_0,
\end{equation}
where $\mathcal U_0 = \mathcal U|_{\mathcal I^+}$ is a radially-independent
contribution fixed by the worldtube boundary conditions. $\mathcal U$ obeys the same radial
differential equation as $\breve U$, but possesses a constant asymptotic value
that is used to determine the evolution of the angular coordinates.

The computational procedure with the gauge transformation to partially flat
coordinates is then:
\begin{enumerate}
\item Perform the gauge transformation from the Cauchy gauge metric to the local
Bondi-Sachs coordinates on the worldtube $\Gamma$, generated by geodesics with
null vectors that are outgoing with respect to the worldtube surface.
\item For each spin weighted scalar $I$ in $\{\beta$, $Q$, $U\}$:
  \begin{enumerate}
  \item Transform 
     $I$ to partially flat gauge $\breve I$
    (or $\mathcal U$) via the angular
    coordinates $x^{A}(\breve u, \breve x^{\breve A})$ \footnote{When performing spectral interpolation, we
      require the position of the target collocation points in the source
      coordinate system. See Sec.~\ref{sec:clenshaw} for more details regarding our
      interpolation methods.}. All transformations for these scalars depend only
    on angular Jacobians $\partial_{\breve A} x^{B}$, and are described in Sec.~\ref{sec:worldtube}.
  \item Evaluate the hypersurface equation for the spin-weighted scalar $\breve I$
      using the radial integration methods described in Sec.~\ref{sec:radial_integration}.
    \end{enumerate}
 \item Determine the time derivative of the angular coordinates
   $\partial_{\breve u} x^{A}(\breve x)$ (see Sec.~\ref{sec:worldtube}) using
   the asymptotic value of $\mathcal U$.
 \item Transform $\mathcal U$ to the partially flat gauge $\breve U$ by
   subtracting its asymptotic value $\mathcal U_0 \equiv \mathcal U|_{\mathcal I^+}$.
 \item For each spin weighted scalar $I$ in $\{W, H\}$:
   \begin{enumerate}
   \item Transform
     $I$ to partially flat gauge $\breve I$ via the
     angular coordinates $ x^A(\breve x^{\breve A})$ and their first derivatives
     $\partial_{\breve u} x^A(\breve x)$ -- see Sec.~\ref{sec:worldtube}.
   \item Evaluate the hypersurface equation for $\breve I$.
   \end{enumerate}
   \item For each output waveform quantity $O$ in $\{h, N, \Psi_4, \Psi_3, \Psi_2,
     \Psi_1, \Psi_0\}$:
     \begin{enumerate}
       \item Compute asymptotic value of $O$, and transform to asymptotically
       inertial coordinate time as described in App.~\ref{app:scri_interpolation},
       using $\mathring{u}(\breve x^{\breve A})$.
     \end{enumerate}
   \item Step $\breve J$ forward in time
     using $\partial_{\breve u} \breve J=\breve H$,
     step $x^A$ using Eq.~(\ref{eq:du_angular_coords}) below for
     $\partial_{\breve u} x^A$, and step
     $\mathring{u}$ using Eq.~(\ref{eq:du_inertial_retarded_time})
       below for $\partial_{\breve u} \mathring{u}$.
\end{enumerate}
See Sec.~\ref{sec:clenshaw} for details regarding the
calculation of the angular Jacobian factors required for the gauge
transformation and the practical methods used to evolve the angular coordinates.

\subsection{Worldtube data interpolation and transformation} \label{sec:worldtube}

The collection of hypersurface equations \eqref{eq:hierarchy} requires data for
each of the quantities $\{\breve \beta,\breve Q,\breve U,\breve W,\breve H\}$ on
a single spherical shell at each timestep.
For $\breve\beta$ and $\breve U$, the worldtube data specifies the
constant-in-$\breve y$ part of the solution on the hypersurface, for $\breve Q$
and $\breve W$, the worldtube data fixes the $\propto (1-\breve y)^2$ part, and
for $\breve H$, the worldtube data fixes a combination of radial modes that
includes the $\propto (1-\breve y)$ contribution.

The worldtube data provided by a Cauchy simulation contains the spacetime
metric, as well as its first radial and time derivatives. The procedure for
transforming the data provided by the Cauchy evolution to boundary data for the
hypersurface equations \eqref{eq:hierarchy} is then, for each hypersurface time
$\breve u$,
\begin{enumerate}
\item Interpolate the worldtube data to the desired hypersurface time $\breve u$
\item Perform the local transformation of the Cauchy worldtube metric and its derivatives to a Bondi-like gauge
  as described in \cite{Bishop:1998uk}
\item Perform angular transformation and interpolation from the generic
  Bondi-like gauge to the partially flat gauge used for the evolution quantities.
\end{enumerate}

The worldtube data is usually generated by the Cauchy simulation at time steps
that are suited to the strong-field calculations, but
the characteristic system can usually
take significantly larger time steps.
Once the characteristic time stepping infrastructure has selected a desired time step, we
interpolate the worldtube data at each angular collocation point to the target
time for the next hypersurface.
In SpECTRE, the interpolation is performed by selecting a number of time points
as centered as possible on the target time, then performing a barycentric
rational interpolation to the target time.

After the time interpolation of the worldtube data, we have the values of
the spacetime metric and its radial and time derivatives on a single inner
boundary of the CCE hypersurface of constant retarded time $\breve u$.
We then compute the outgoing radial null vector $l^{\mu'}$ (denoting
Cauchy coordinate quantities with $'$) 
, construct a radial null
coordinate system using the affine parameter along null geodesics generated by
$l^{\mu'}$, then normalize the radial coordinate to construct an areal
radius $r$.
Following these transformations, for which explicit formulas are given in
\cite{Bishop:1998uk, Barkett:2019uae, Moxon:2020gha}, the spacetime metric
$g_{\alpha \beta}$ is of the form \eqref{eq:bs_metric}, but with no
asymptotic flatness behavior imposed.  
During the transformation from the Cauchy coordinates to the Bondi-like
coordinates, the angular and time coordinates remain fixed on the worldtube
surface, so no alteration of the pseudospectral grid is necessary.

The final step for the worldtube computation is to perform a constant-in-$r$
angular coordinate transformation to a set of angular coordinates $x^A(\breve
x^{\breve A})$ for which the metric satisfies the asymptotic conditions:
\begin{subequations}
  \begin{align}
    \lim_{\breve y \rightarrow 1} \breve J = 0,\\
    \lim_{\breve y \rightarrow 1} \breve U = 0.
  \end{align}
\end{subequations}

These conditions are satisfied if the angular coordinates obey the
radially-independent evolution equation \cite{Moxon:2020gha}
\begin{equation} \label{eq:du_angular_coords}
  \partial_{\breve u} x^{A} = - \mathcal U_0^{\breve A} \partial_{\breve A} x^A,
\end{equation}
where $\mathcal U_0^{\breve A} \breve q_{\breve A} \equiv \mathcal U_0$.

The angular transformations for the remaining spin-weighted scalars require the
spin-weighted angular Jacobian factors
\begin{subequations}\label{eq:angular_jacobians}
  \begin{align}
    \breve a = \breve q^{\breve A} \partial_{\breve A} x^B q_B\\
    \breve b= \breve{\bar q}^{\breve A} \partial_{\breve A} x^B q_B\\
  \end{align}
\end{subequations}
and conformal factor
\begin{subequations}
  \begin{align}
    \breve \omega &= \frac{1}{2}\sqrt{\breve b \breve {\bar b} - \breve a \breve{\bar a}}\\
    \partial_{\breve u} \breve \omega &= \frac{\breve \omega}{4}\left(\breve{\bar \eth} \mathcal U_0 + \breve \eth \bar{\mathcal U}_0\right) + \frac{1}{2} \left(\mathcal U_0 \breve{\bar \eth} \breve \omega + \bar{\mathcal U}_0 \breve \eth \breve \omega\right)
  \end{align}
\end{subequations}

Given the angular coordinates determined by the time evolution of
\eqref{eq:du_angular_coords}, we perform interpolation of each of the
spin-weighted scalars $\{R, \partial_{u} R, J, U, \partial_r U, \beta,
Q, W, H\}$ to the new angular collocation points (more details for the
numerical interpolation procedure are in Sec.~\ref{sec:clenshaw}), and
perform the transformation of the spin-weighted scalars as
\begin{widetext}
{\allowdisplaybreaks
  \begin{subequations} \label{eq:partially_flat_tensor_transformations}
  \begin{align}
    \breve R =& \breve \omega R, \\
    \partial_{\breve u} \breve R =& \breve \omega \partial_u R + \partial_{\breve u} \breve \omega + \frac{\breve \omega}{2}\left(\mathcal U_0 \breve{\bar \eth} R + \bar {\mathcal U}_0 \breve \eth R\right),\\
    \breve J =& \frac{1}{4 \breve \omega^2} \left(\breve{\bar b}^2 J + \breve a^2 \bar J + 2 \breve a \breve{\bar b} K\right),\\
    e^{2 \breve \beta} =& \frac{e^{2 \beta}}{\breve \omega},\\
    \partial_{\breve y} \breve U =& \frac{\breve R}{\breve \omega^3 (1 - \breve y)^2} \left(\breve{\bar b} \partial_r U - \breve c \partial_r \bar U\right)
                                    + 4 \breve R\frac{e^{2 \breve \beta}}{\breve \omega}\left[\breve{\bar \eth} \breve \omega \partial_{\breve y} \breve J - \breve \eth \breve \omega \left(\frac{\partial_{\breve y}(\breve J \breve{\bar J})}{2 \breve K}\right)\right]\notag\\
              &+ 2 \breve R \frac{e^{2 \breve \beta}}{\breve \omega}\left(\breve J \breve{\bar \eth} \breve \omega - \breve K \breve \eth \breve \omega\right)\left[-1 + \partial_{\breve y} \breve{\bar J} \partial_{\breve y} \breve J - \left(\frac{\partial_{\breve y}(\breve J \breve{\bar J})}{2 \breve K}\right)^2\right],\\
    \breve Q =& 2 \breve R e^{-2 \breve \beta} \left(\breve K \partial_{\breve y} \breve U + \breve J \partial_{\breve y} \breve{\bar U}\right),\\
    \mathcal U =& \frac{1}{2 \breve \omega} (\breve{\bar b} U - \breve c \bar U) - \frac{e^{2 \breve \beta} (1 - \breve y)}{2 \breve R \breve \omega}(\breve K \breve\eth \breve \omega - \breve J \breve{\bar \eth} \breve \omega),\\
    \breve U =& \mathcal U - \mathcal U_0,\label{eq:u_gauge_change}\\
    \breve W =& W + \frac{(\breve \omega - 1)(1 - \breve y)}{2 \breve R} + \frac{e^{2 \breve \beta}(1 - \breve y)}{4 \breve R \breve \omega^2}\left[\breve J (\breve{\bar \eth} \breve \omega)^2 + \breve{\bar J}(\breve \eth \breve \omega)^2 - 2 \breve K (\breve \eth \breve \omega)(\breve{\bar \eth} \breve \omega)\right] - \frac{2 \partial_{\breve u} \breve \omega}{\breve \omega} - \frac{\breve U \breve{\bar \eth} \breve \omega + \breve{\bar U} \breve \eth \breve \omega}{\breve \omega},\\
    \breve H =& \frac{1}{2}\left[\mathcal U_0 \breve{\bar \eth} \breve J + \breve \eth (\bar{\mathcal U}_0  \breve J) - \breve J\breve \eth \bar{\mathcal U}_0 \right] + \frac{\partial_{\breve u} \breve \omega - \frac{1}{2}(\mathcal U_0 \breve{\bar \eth} \breve \omega + \bar{\mathcal U}_0 \breve \eth \breve \omega)}{\breve \omega}(2 \breve J - 2 \partial_{\breve y} \breve J) - \breve J \breve{\bar \eth} \mathcal U_0 + \breve K \breve \eth \bar {\mathcal U}_0 \notag\\
              & \frac{1}{4 \breve \omega} \left(\breve{\bar b}^2 H + \breve a^2 \bar H + \breve{\bar b} \breve c \frac{H \bar J + J \bar H}{K}\right) + 2 \frac{\partial_{\breve u} \breve R}{\breve R} \partial_{\breve y} \breve J,
  \end{align}
\end{subequations}
}
\end{widetext}
where $K = \sqrt{1 + J \bar J}$ and $\breve K = \sqrt{1 + \breve J \breve{\bar J}}$.
Finally, the quantities $\{ \breve \beta, \breve Q,
\mathcal U, \breve W, \breve H\}$ are used directly to determine the integration
constants in the hypersurface equations \eqref{eq:hierarchy}.
Note that in all of the equations \eqref{eq:u_gauge_change} onward, we have
explicit dependence on $\mathcal U_0$ or implicit dependence on $\mathcal U_0$
via $\partial_{\breve u} \breve \omega$. This dependence necessitates
finishing the hypersurface integration of $\mathcal U$ to determine its
asymptotic value before computing the remaining gauge-transformed quantities on
the worldtube.

\subsection{Initial data} \label{sec:initial_data}

In addition to the specification of the worldtube data at the interface to the
Cauchy simulation, the characteristic
system requires initial data at the first outgoing
null hypersurface in the evolution (see Fig.~\ref{fig:cce_sketch}).
The initial data problem on this hypersurface is physically similar to the
initial data problem for the Cauchy evolution: It is computationally prohibitive
to directly construct the spacetime metric in the state that it would possess
during the inspiral.
Ideally, we would like the starting state of the simulation to be simply
a snapshot of
the state if we had been simulating the system for far longer.

The initial data problem in CCE has been investigated previously by
\cite{Bishop:2011iu}, in which a linearized solution scheme was considered.
The most important part of the initial data specification appears to be choosing the
first hypersurface such that it is consistent with the boundary data at the same
timestep.
Without that constraint, previous authors \cite{Bishop:2011iu}, and empirical
tests of our own code, indicate that spurious oscillations emerge that often
last the full duration of the simulation.

Computationally, the initial data freedom in CCE is 
much simpler than the Cauchy case \cite{York:1998hy,Pfeiffer:2002iy}.
We may specify the Bondi-Sachs transverse-traceless angular scalar $\breve J$
arbitrarily.
Even when we take the practical constraint that $\breve J$
must be consistent with the
worldtube data at the first timestep, we still have almost arbitrary freedom in
the specification of $J$, as it must be consistent with the worldtube data only
up to an arbitrary angular coordinate transformation \footnote{In our evolution
  system, we track and perform an angular coordinate transformation at the
  worldtube regardless of initial data choice, so permitting this transformation
  on the initial hypersurface amounts only to setting nontrivial initial data
  for $x^A(\hat x^{\hat A})$.
}.

\begin{figure}
  \includegraphics[width=.48\textwidth]{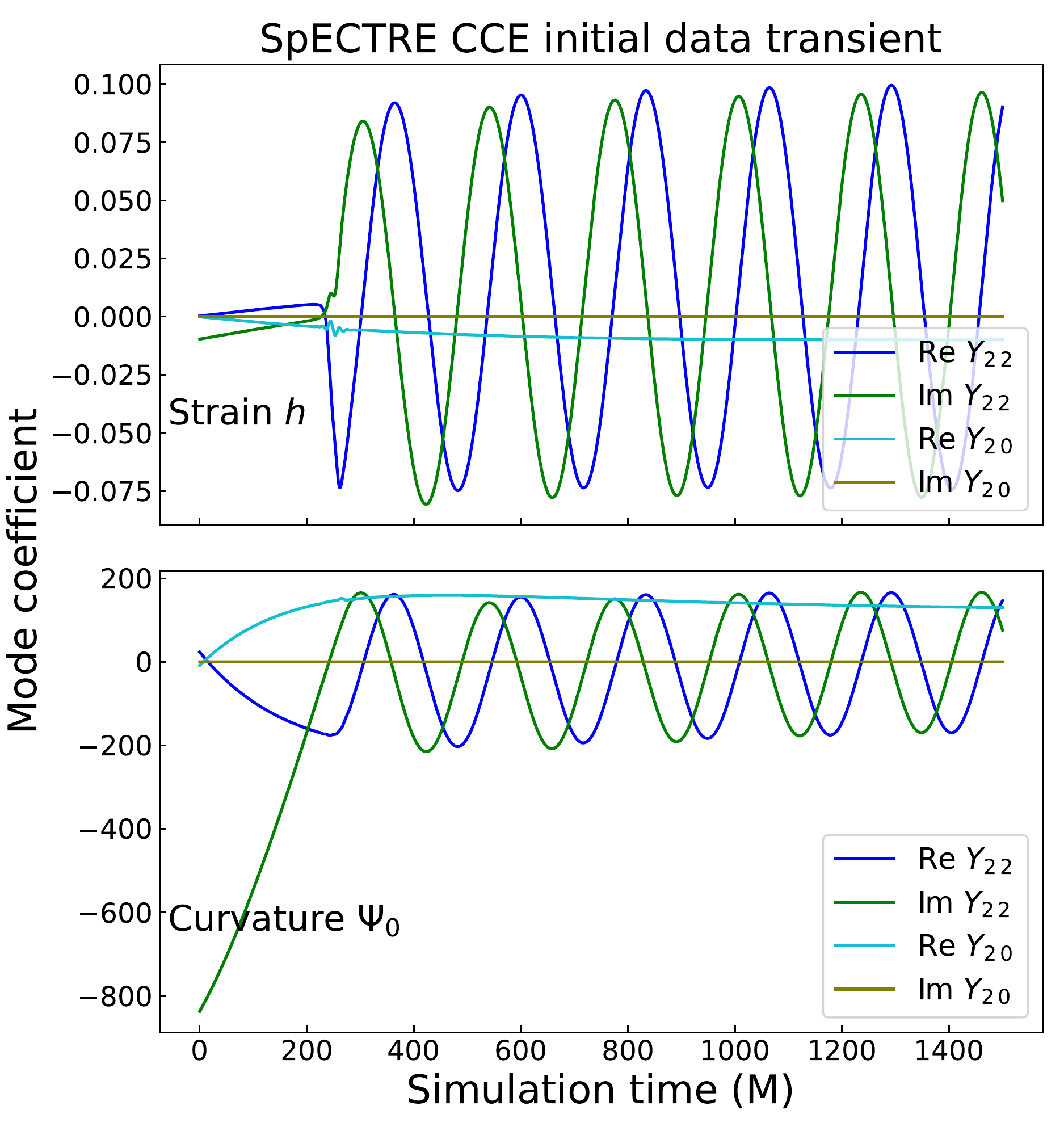}
  \caption{The initial data transient for an example CCE run using worldtube
    data obtained from a binary black hole simulation
    \texttt{SXS:BBH:2096} from the
    SXS catalog. The dominant modes of the strain and $\Psi_0$ display visually
    apparent drift during the first $\sim 2$ orbits of the inspiral. The initial
    data transient contaminates the data for the early part of the simulation and
    leads to a BMS frame shift in the strain waveform.  The frame shift can be
    seen visually from the fact that the $Y_{2 2}$ mode does not oscillate about 0.
    The initial data method used for this demonstration is the cubic ansatz initial
    data described as method \ref{cubic_id} below.
  }
  \label{fig:id_transient}
\end{figure}

Current methods of choosing initial data for $J$ do not represent a snapshot of
a much longer simulation, and this gives rise to transients in the resulting
strain outputs (see Fig.~\ref{fig:id_transient}).
These initial data transients are analogous to `junk radiation' frequently found
in Cauchy simulations, but are somewhat more frustrating for data analysis
because the CCE initial data transients tend to have comparatively long
timescales.
We observe that the strain waveform 
tends to settle to a suitable state within a few
orbits of the start of the simulation.
However, when recovering high-fidelity waveforms from an expensive Cauchy
simulation, every orbit of trustworthy worldtube data is precious, and it is
disappointing to lose those first orbits of data to the initial data transient.
It is a topic of ongoing work to develop methods of efficiently generating
high-quality initial data for CCE to improve the initial data transient
behavior (see Sec. \ref{sec:improving_initial_data}).

We currently support three methods for generating initial hypersurface data:\\
\begin{enumerate}
\item Keep $\breve J$ and $\partial_{\breve y} \breve J$ consistent with the
first timestep of the worldtube data. Use those quantities to fix the
angularly dependent
coefficients $A$ and $B$ in the cubic initial hypersurface ansatz:
\begin{equation}
  \breve J(\breve y, \breve \theta, \breve \phi) = A(\breve \theta, \breve \phi) (1- \breve y) + B(\breve \theta, \breve \phi) (1 - \breve y)^3.
\end{equation}
This is a similar initial data construction to \cite{Bishop:2011iu}, and is
chosen to omit any $(1-\breve y)^2$ dependence, which guarantees that no
pure-gauge logarithmic terms arise during the evolution \cite{Moxon:2020gha}.\label{cubic_id}\\
\item Set  the Newman-Penrose quantity $\Psi_0 = 0$ on the initial hypersurface.
  This amounts to enforcing
a second-order nonlinear ordinary differential equation in $y \equiv 1 - 2 R/ r$ for
$J$, before constructing the coordinate
  transformation from $x^{\alpha}$ to $\breve x^{\breve \alpha}$.
After some simplification, the expression for $\Psi_0$ in
\cite{Moxon:2020gha} may be used to show that the equation
\begin{align} \label{eq:psi_0_radial_ode}
\partial_{ y}^2  J = \frac{1}{16  K^2} &\left({\bar J}^2 (\partial_{ y}  J)^2 - 2(2 +  J {\bar J}) \partial_{ y}  J \partial_{ y} {\bar J} +  J^2 (\partial_{ y} {\bar J})^2\right)\notag\\&\times\left(-4  J -(1 -   y) \partial_{ y}  J\right)
\end{align}
is equivalent to the condition $\Psi_0 = 0$.
The initial hypersurface data is generated by first using
(\ref{eq:psi_0_radial_ode}) to perform a radial ODE integration out to $\mathcal
I^+$, with boundary values of $J$ and $\partial_{y} J$ on the initial worldtube.
However, the data so generated is not necessarily asymptotically flat, so an
angular coordinate transformation is calculated to fix $\breve J|_{\mathcal I^+}
= 0$.
Encouragingly, fixing both \eqref{eq:psi_0_radial_ode} and the asymptotic
flatness condition also constrains the $(1 - y)^2$ part of $J$
to vanish, which is sufficient to
prevent the emergence of pure-gauge logarithmic dependence during the evolution
of $J$. \label{no_incoming_radiation_id}\\
\item Set $\breve J=0$ along the entire initial hypersurface. In general,
this choice will be inconsistent with the data specified on the worldtube
$J|_{\Gamma}$, so it is necessary to construct an angular transformation $
x(\breve x^{\breve A})$ such that $\breve J|_{\Gamma} = 0$ following the
transformation. \label{zero_nonflat_id}
\end{enumerate}
Methods \ref{no_incoming_radiation_id} and \ref{zero_nonflat_id} above require the ability to compute the angular
coordinate transformation $x^A(\hat x^{\hat B})$ such that
\begin{equation} \label{eq:angular_coordinate_constraint}
0 = \breve J = \frac{\breve{\bar b}^2 \breve J + \breve a^2 \breve{\bar J} + 2 \breve a \breve{\bar b} \breve K}{4 \breve \omega^2}
\end{equation}
on some surface.
Solving \eqref{eq:angular_coordinate_constraint} in general would amount to an
expensive high-dimensional root-find.

However, in our present application, practical solutions in the wave zone
typically have a value of $\breve J$ no greater than $\sim 5 \times 10^{-3}$,
and we should not expect to find a well-behaved angular coordinate transform
otherwise.
So, we take advantage of the small parameter in the equation to iteratively
construct candidate angular coordinate systems that approach the condition
\eqref{eq:angular_coordinate_constraint}.
Our linearized iteration is based on the approximation
\begin{subequations}
\begin{align}
  \breve a_{n+1} &= -\frac{1}{2} \frac{\breve J_{n} \breve{\omega}_{n}}
  {\breve{\bar b}_n \breve K_n}\\
  \breve x^{i}_{n+1}(\breve x) &= \frac{1}{2}\breve \eth^{-1}_{n+1}
  \left(\breve a_{n + 1} \breve \eth \breve x^i + \breve{\bar b}_{n+1} \breve{\bar \eth} \breve x^i\right),
\end{align}
\end{subequations}
for a collection of Cartesian coordinates $\breve x^i$ that are representative of
the angular coordinate transformation (see Sec.~\ref{sec:clenshaw}).

We find that this procedure typically approaches roundoff in $\sim 10^3$
iterations.
Despite the crude inefficiency of this approximation, the iterative solve needs
to be conducted only once, so it represents only a small portion of the CCE
execution time for the initial data methods that take advantage of it.

In practical investigations, it has been found that most frequently the simplest
method of an inverse cubic ansatz ({\bf 1.} above) performs best in various
measures of asymptotic data quality \cite{Mitman:2021xkq}. However, because the
reasons for the difference in precision for different initial data schemes are
not currently well understood, we believe it useful to include descriptions of
all viable methods.

\section{Implementation details and Numerical optimizations} \label{sec:numerical_optimizations}

Much of the good performance of the SpECTRE CCE system is inherited
from the shared SpECTRE infrastructure.
In particular, the SpECTRE data structures offer easy interfaces to aggregated
allocations (which limit expensive allocation of memory), fast vector operations
through the interface with the open source {\sc Blaze} library \cite{Blaze}, and
rapid SWSH transforms via the open source {\sc libsharp} library.
Further, we take advantage of per-core caching mechanisms to avoid recomputing
common numerical constants, such as spectral weights and collocation values.

However, in addition to establishing ambitious ``best practices'' for the
mechanical details of the software development, we have implemented numerical
optimizations specialized to calculations in the CCE system.
We give a brief explanation of the techniques we use to improve performance of
angular interpolation in Sec.~\ref{sec:clenshaw}, which is required to perform
the gauge transformation discussed in Sec.~\ref{sec:worldtube}. In
Sec.~\ref{sec:radial_integration}, we explain our methods for efficiently
performing the hypersurface integrals in our chosen Legendre-Gauss-Lobatto
pseudospectral representation.

\subsection{Angular interpolation techniques using spin-weighted Clenshaw algorithm} \label{sec:clenshaw}

The Clenshaw recurrence algorithm is a fast method of computing the sum over basis functions,
\begin{equation} \label{eq:BasisSum}
  f(x) = \sum_{n = 0}^{N} a_n \phi_n(x),
\end{equation}
provided the set of basis functions $\phi_n$ obeys a standard form of a
three-term recurrence relation common to many polynomial bases.
In particular, it is assumed that $\phi_n$ may be written as,
\begin{equation}
  \phi_n(x) = \alpha_n(x) \phi_{n - 1}(x) + \beta_{n}(x) \phi_{n - 2}(x),
\end{equation}
for some set of easily computed $\alpha_n$ and $\beta_n$.

The algorithm for computing the full sum $f(x)$ \cite{Press:2007} is then to
compute the set of quantities $y_n$ for $n \ge 1$, where $y_n$ is
\begin{subequations}
\begin{align}
  y_{N+2}(x) =& y_{N+1}(x) = 0\\
  y_{n}(x) =& \alpha_{n + 1}(x) y_{n + 1}(x) \notag\\
             &+ \beta_{n + 2}(x) y_{n + 2}(x) + a_n
\end{align}
\end{subequations}
Once the last two quantities in the chain $y_1(x)$ and $y_2(x)$ are determined,
the final sum is obtained from the formula
\begin{equation}
  f(x) = \beta_2(x) \phi_0(x) y_2(x) + \phi_1(x) y_1(x) + a_0 \phi_0(x).
\end{equation}

We use the Clenshaw method for interpolating SWSH data to arbitrary points $x$
on the sphere.
For spherical harmonics, it is successive values of $\ell$ that have convenient
three-term recurrence relations, so the lowest modes in the recursion are
$Y_{|m|, m}(\theta, \phi)$ and $Y_{|m|+1, m}(\theta, \phi)$.
The values of $\alpha_{\ell, m}(\theta, \phi)$ and $\beta_{\ell, m}(\theta, \phi)$ are
cached for the target interpolation points, and the source collocation
 values are
transformed to spectral coefficients $a_{\ell, m}$.
The Clenshaw algorithm can be applied directly at
each of the target
points $(\theta, \phi)$, to obtain the values $f(\theta, \phi)$.
Note that the step of caching the $\alpha_{\ell,m}(\theta, \phi)$ and
$\beta_{\ell,m}(\theta, \phi)$ is primarily useful for interpolating multiple
functions to the same grid; if only one function is needed for each grid, there
will be little gain in caching $\alpha$ and $\beta$, as they would each be
evaluated only once in a given recurrence chain.

In Appendix~\ref{sec:clenshaw_details}, we give full details of the specific
recurrence relations that can be used to efficiently calculate the Clenshaw sum
for SWSH, as well as additional recurrence relations that improve performance
when moving between the $m$ modes.
For the remaining discussion it is convenient to define a few auxiliary
variables that are used in the formulas for the SWSH recurrence:
\begin{subequations} \label{eq:abValues}
  \begin{align}
    a &= |s + m| \\
    b &= |s - m| \\
    \lambda &= \begin{cases}0, &  s \ge -m\\
      s + m, &  s < -m \end{cases}
  \end{align}
\end{subequations}

The step-by-step procedure for efficiently interpolating a spin-weighted
function represented as a series of spin-weighted spherical harmonic
coefficients to a set of target collocation points $(\theta_i, \phi_i)$ is then:
\begin{enumerate}
\item Assemble the lookup table of required ($\alpha_{\ell}^{(a, b)}(\theta)$, $\beta_{\ell}^{(a, b)}$, $\lambda_m$):
  \begin{enumerate}
 \item For each $m \in [-\ell_{\max}, \ell_{\max}]$ there is a pair $(a, b)$ from
    (\ref{eq:abValues}) to be computed.
    Note that $\alpha^{(a, b)}_\ell$ must be cached separately for each target point, but
    $\beta^{(a, b)}_\ell$ does not depend on the target coordinates.
  \end{enumerate}
\item For $m \in [0, \ell_{\max}]$:
  \begin{enumerate}
 \item If $|s| \ge |m|$: Determine ${}_s Y_{|s|, m}(\theta, \phi)$ from direct
    evaluation of (\ref{eq:SwshToWigner}) with (\ref{eq:WignerToJacobi}) and
    ${}_s Y_{|s| + 1, m}(\theta, \phi)$ from (\ref{eq:SecondL}); Store
    ${}_s Y_{|s|, m}(\theta, \phi)$ for recursion if $|s| = |m|$.
  \item If $|m| > |s|$: Determine ${}_s Y_{|m|, m}(\theta, \phi)$ from
    recurrence (\ref{eq:MRecursion}) and ${}_s Y_{|m| + 1, m}(\theta, \phi)$
    from (\ref{eq:SecondL}).
    Store ${}_s Y_{|m|, m}(\theta, \phi)$ for recursion.
  \item Perform the Clenshaw algorithm to sum over $l \in [\min(|s|, |m|),
    \ell_{\max}]$, using the spectral coefficients $a_{\ell m}$, the precomputed
    $\alpha_{\ell}^{(a, b)}$ and $\beta_{\ell}^{(a, b)}$ recurrence coefficients, and the first
    two harmonics in the sequence computed from the previous step.
  \end{enumerate}
\item For $m \in [-1, -\ell_{\max}]$, repeat the substeps of step 2, but for
  the negative set of $m$'s.
\end{enumerate}

Although the procedure for interpolation is performed efficiently, there are a
number of details of the implementation of the angular coordinate transformation
that must be handled carefully.

\begin{figure}
  \includegraphics[width=.36\textwidth]{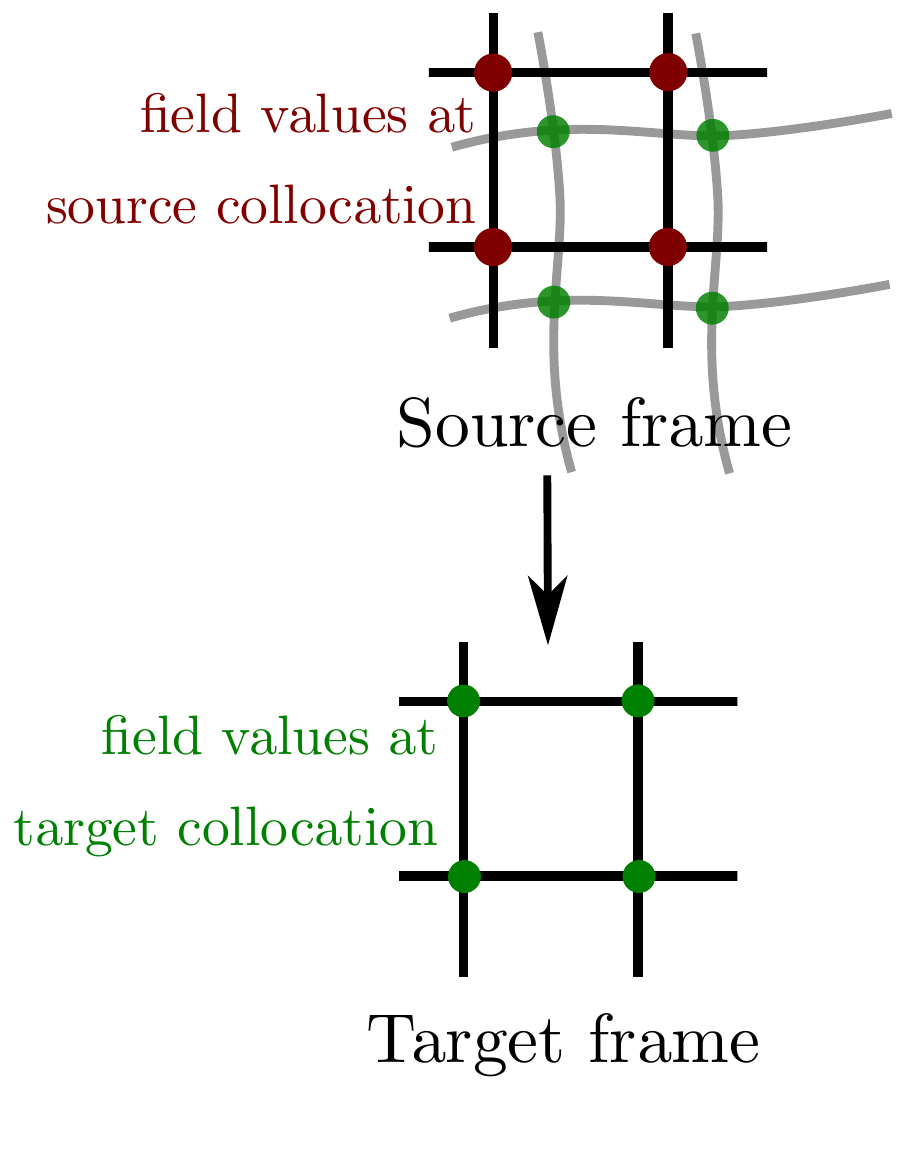}
  \caption{An illustration of the interpolation reasoning for pseudospectral methods.
  The input to the interpolation is the field values at the collocation points
in the source frame, and we wish to determine the field values for the same
function at the collocation points in the target frame, which will be at
non-collocation points in the source frame coordinates.
Therefore, the interpolation seeks to calculate the field value at points
$x(\hat x)$ in the source frame, for all collocation points $\hat x$ in the
target frame. }
  \label{fig:interpolation}
\end{figure}

First, it is important to note the counterintuitive nature of the set of
coordinate functions we require for the interpolation.
In both the source frame and the target frame, we use a pseudospectral grid,
evenly spaced in $\phi$, and at Legendre-Gauss points in $\theta$.
When interpolating, we require the location in the source frame coordinates of
the target frame collocation points.
Therefore, when expressed as a function over collocation points, the function
that we use for interpolation is $x^A(\hat x^A)$.  
We have found this feature of the interpolation for pseudospectral methods
easy to misremember, so we have included Fig.~\ref{fig:interpolation} to
assist in recalling the correct reasoning.

Most of the quantities that we wish to interpolate have nonzero spin-weight,
so do not transform as scalars.
Instead, their transformation involves factors of the spin-weighted angular
Jacobians \eqref{eq:angular_jacobians}. The tensor transformations for each of the
relevant quantities at the worldtube boundary are given in
\eqref{eq:partially_flat_tensor_transformations}.
For illustration, let us discuss the transformation of the spin-weight 2 scalar
$\breve J$:
\begin{equation} \label{eq:J_transform}
  \breve J = \frac{\breve{\bar b}^2  J + \breve a^2 \bar J + 2 \breve a \breve{\bar b}  K}{4 \breve \omega}
\end{equation}
It is important to note that at the start of the transformation procedure, we
have the values of $ J$ on the source grid $x^A$ and the values of $\breve a$,
$\breve b$, and $\breve \omega$ on the target grid $\breve x^{\breve A}$ (the
Jacobians are derivatives of $x(\breve x)$; see Fig.~\ref{fig:interpolation}).

The spin-weighted interpolation procedure can be performed only on quantities
that are representable by the SWSH basis.   
We can store non-representable quantities (including, e.g. the angular coordinates
themselves) on our chosen angular grid,
but we cannot perform a SWSH transform on such quantities, so we
cannot interpolate them using pseudospectral methods with any predictable
accuracy.
Inconveniently, we are burdened with a number of quantities that are not
representable on the SWSH basis.
Immediately after interpolation, $J(x^A(\breve x^{\breve A}))$ is not
representable on the basis corresponding to the new grid because the Jacobian
factors have not yet been applied.
Similarly, the Jacobian factors $\breve a$ and $\breve b$ are not representable
on the SWSH basis whenever the angular transform is not trivial.

Accordingly, for our example of $\breve J$, we must apply the transformation
operations in a specific sequence:
\begin{enumerate}
\item Interpolate $J(x^A)$ and $K(x^A)$ to $J(x^A(\breve x^{\breve A}))$ and
  $K(x^A(\breve x^{\breve A}))$.
\item Multiply the result by the Jacobian factors that appear in
  \eqref{eq:J_transform}.
\end{enumerate}

We meet a similar complication when manipulating the evolved angular coordinates
$x^A(\breve u, \breve x^{\breve A})$. The angular coordinates are not
representable on the SWSH basis, yet we must take angular derivatives of the
angular coordinates to determine the Jacobian factors
\eqref{eq:angular_jacobians}.
The method we use to evade the problems for the angular coordinate
representation is to introduce a unit sphere Cartesian representation of the
angular coordinates:
\begin{subequations}
  \begin{align}
    x_{\text{unit}} &= \sin\theta \cos\phi,\\
    y_{\text{unit}} &= \sin\theta \sin\phi,\\
    z_{\text{unit}} &= \cos\theta.
  \end{align}
\end{subequations}
The evolution equation for the unit sphere Cartesian representation is then
derived from the angular coordinate evolution equation
\eqref{eq:du_angular_coords}.
\begin{align}
  \partial_{\breve u} x^i_{\text{unit}} &= \mathcal U_0^{\breve A} \partial_{\breve A} x^i_{\text{unit}}\notag\\
                                        &=\frac{1}{2}\left(\mathcal U_0 \breve{\bar\eth} x^i_{\text{unit}} + \bar{\mathcal U}_0 \breve{\eth} x^i_{\text{unit}}\right).
\end{align}

The main advantage of promoting the angular coordinates $x^A(\breve u, \breve
x^{\breve A})$ to their unit sphere Cartesian analogs is that the Cartesian
coordinates $x^i$ are spin-weight 0 and so we can quickly and accurately
evaluate their angular derivatives.

The spin-weighted Jacobian factors \eqref{eq:angular_jacobians} are then calculated as
\begin{subequations}
  \begin{align}
    \breve a = \breve \eth x^i \partial_i x^A q_A,\\
    \breve b = \breve{\bar \eth} x^i \partial_i x^A q_A,
  \end{align}
\end{subequations}
where the factors $\partial_i x^A$ are the Cartesian-to-angular Jacobians in the
source frame, so are analytically computed as
\begin{subequations}
  \begin{align}
    \partial_x \theta &= \cos[\phi(\hat x^{\hat A})] \cos[\theta(\hat x^{\hat A})], \\
    \partial_x \phi &= - \sin[\phi(\hat x^{\hat A})] / \sin[\theta(\hat x^{\hat A})],\\
    \partial_y \theta &= \cos[\theta(\hat x^{\hat A})] \sin[\phi(\hat x^{\hat A})] \\
    \partial_y \phi &= \cos[\phi(\hat x^{\hat A})] / \sin[\theta(\hat x^{\hat A})],\\
    \partial_z \theta &= -\sin[\theta(\hat x^{\hat A})],\\ \partial_z \phi &= 0.
  \end{align}
\end{subequations}

\subsection{Rapid linear algebra methods for radial integration} \label{sec:radial_integration}

SpECTRE CCE uses a Legendre Gauss-Lobatto spectral representation for the radial
dependence of the spin-weighted scalars on its domain.
The use of spectral methods  allows rapid integration of the radial differential equations of the
hierarchical CCE system \eqref{eq:hierarchy}.
The numerical methods we employ in this section are not themselves new, but they
have not previously been applied to efficiently solving the CCE system of
equations.

Each of the angular derivatives that appears in the hierarchy of radial
differential equations is first evaluated by the procedure described
around Eq.~\eqref{eq:spin_weighted_derivative_factors}:
perform a spin-weighted
spherical harmonic transform using {\sc libsharp}, multiply by $\sqrt{(\ell -
s)(\ell + s + 1)}$ in the modal basis for the $\breve\eth$ and $-\sqrt{(\ell + s)(\ell -
s + 1)}$ for $\breve{\bar \eth}$, and recover the nodal representation of the
derivative with an inverse spin-weighted transform.
Using these
nodal values of the angular derivative terms ,
we may then directly compute each of the right-hand sides of the
radial differential equations over the nodal grid.
Therefore, for each of the radial differential equations, the problem reduces to
a collection of radial ODE solves.

The spectral representation in the radial direction allows the further
simplification of determining linear operators that correspond to indefinite
integration. Given the function
$f$ expressed in the modal representation
\begin{equation} \label{eq:modal_representation}
  f(\breve y) = \sum_n a_n P_n(\breve y),
\end{equation}
we seek the integration matrix $I$ such that 
\begin{align} \label{eq:integration_matrix}
  &\sum_n a_n \int^{\breve y} P_n(\breve y) = \sum_n (I \cdot a)_n  P_n(\breve y),\notag\\
  \Longrightarrow\quad & \sum_n a_n P_n(\breve y) = \sum_n (I \cdot a)_n \partial_{\breve y} P_n(\breve y),
\end{align}
The relevant identity for Legendre polynomials that we use to determine the
integration matrix $I$ is
\begin{equation} \label{eq:GL_differentiation_identity}
  P_n(\breve y) = \frac{1}{2n + 1} \frac{d}{d\breve y}\left[P_{n + 1}(\breve y) - P_{n - 1}(\breve y)\right].
\end{equation}
By integrating both sides of this equation and applying the
result to the modal representation \eqref{eq:modal_representation}, we find the
almost-tridiagonal indefinite integration matrix for the spectral representation
\begin{equation} \label{eq:integration_matrix_matrix_rep}
  I =
  \begin{bmatrix}
    -1 & 1 & -1 & 1 & \cdots& (-1)^{n+1}\\
    -1 & 0 & -1/3 & 0& \cdots& 0\\
    0 & 1 & 0 & -1/5 & \cdots& 0\\
    \vdots & \vdots & \ddots & \ddots &\ddots & \vdots \\
     0 & 0 & \cdots&  1/(2n - 1) & 0 & -1/(2n + 3)
  \end{bmatrix}.
\end{equation}
Here the first row is chosen to zero the function at the innermost gridpoint
(at $\breve y = -1$). It is convenient
to generate linear operators acting entirely on the
nodal representation. These are composed as $M^{-1} I M$, where $M$ is the
linear operator that maps the nodal representation to the modal representation.
We may then add an integration constant freely to the result of the indefinite
integration operator in the nodal representation to satisfy the boundary
conditions.

Two of the five equations (those that determine $\breve \beta$ and $\breve U$)
take the simple form
\begin{equation}
  \partial_{\breve y} f = S_f.
\end{equation}
The radial ODE solves for these cases are a straightforward application of the
nodal integration matrix $M^{-1} I M$ using
\eqref{eq:integration_matrix_matrix_rep}.    
In the CCE system, the choice to zero the value at the
innermost boundary point ensures that
we may impose the boundary conditions for the worldtube quantities $\breve
\beta|_\Gamma$ and $\breve U|_{\Gamma}$ by adding the appropriate boundary value
to all points along the radial rays for each angular point on the boundary.

Two more of the radial differential equations (those that determine $\breve Q$
and $\breve W$) take the form
\begin{equation} \label{eq:radial_ode_second_kind}
 (1- \breve y) \partial_{\breve y} f + 2 f = S_f.
\end{equation}
This case requires more care than the original indefinite
integral, but the full integration matrix is still readily calculable for
arbitrary Legendre order $n$.

Considering again the modal representation \eqref{eq:modal_representation}, we
wish to find the linear operator $K$ such that
\begin{equation}
  \sum_n a_n P_n(\breve y) = \sum_n (K\cdot a)_n [(1 - \breve y) \partial_{\breve y} P_n(\breve y) + 2 P_n(\breve y)].
\end{equation}
The operator $K$ is the inverse of the operator in Eq. \eqref{eq:radial_ode_second_kind}.

We will again make use of the integration matrix $I$ \eqref{eq:integration_matrix_matrix_rep}.
We also require the inverse of the matrix $C$ associated with multiplication by
$(1 - \breve y)$:
\begin{equation}\label{eq:matrix_c_def}
  \sum (C \cdot a)_n P_n(\breve y) = \sum a_n (1 - \breve y) P_n(\breve y).
\end{equation}
The matrix $C$ is derived by algebraic manipulations of Bonnet's recursion
formula for Legendre polynomials
\begin{align}
  (n + 1) P_{n + 1} &= (2n + 1) \breve y P_n - n P_{n - 1}\notag\\
  \Rightarrow (1 - \breve y) P_n &= - \frac{n + 1}{2n + 1}P_{n + 1} + P_n - \frac{n}{2n + 1} P_{n -1}
\end{align}
Therefore, composing the operations of $C$ and $I$, we find
\begin{equation}\label{eq:compose_c_i}
 \sum_n ((C + 2I)\cdot a)_n P_n(\breve y) = \sum_n (I \cdot a)_n \left[ (1-\breve y) \partial_{\breve y} P_n + 2 P_n\right]
\end{equation}
and
\begin{equation} \label{eq:K_integration_matrix}
  K = I \cdot (C + 2I)^{-1}
\end{equation}
To compute $K$ in practice, we determine the values of $C$ and $I$ analytically,
then perform a single numerical inversion to finish the computation of
\eqref{eq:K_integration_matrix}. Boundary conditions then determine the
quadratic part of the solution, so are imposed by adding the appropriate
$b(\breve \theta, \breve \phi) (1- \breve y)^2$ contribution along each radial ray.

Importantly, for both of the above types of the radial ODE solve, the
integration matrix in question is independent of the
values of the
fields. So, at the start of the simulation, we precompute and store the
necessary integration matrices, reducing each of the ODE solves described above
to a matrix-vector multiplication for each radial ray. In SpECTRE, these
matrix-vector product calculations are optimized via the vector intrinsic
library {\sc libxsmm} \cite{Libxsmm}.

The final type of radial differential equation appears only in the equation that
determines $H$. This type is
more complicated:
\begin{align}
  (1 - \breve y)& \partial_{\breve y} f + [1 + (1 - \breve y) L^G L^J] f + (1- \breve y) \bar{L}^G L^J \bar f = S,
\end{align}
in which the $L$ factors depend on the field quantities of the current
hypersurface.
In this case, there is little hope of determining an elegant simplification
using
the modal basis. In any case, there would be no opportunity for caching and
reusing an integration matrix, as the differential operator that acts on $f$
depends on the other fields on the hypersurface.
So, for the integration of the $H$ equation, we decompose the complex linear
differential equation into a real linear equation on vectors of length $2n$:
\begin{multline} \label{eq:H_eqn_linear_solve}
  \bigg\{
  \begin{bmatrix}
    (1 - \breve y) \partial_{\breve y} + 1 & 0 \\
    0 & (1 - \breve y) \partial_{\breve y} + 1
  \end{bmatrix} \\
        + (1 -\breve y)
        \begin{bmatrix}
          \text{Re}(L^J) \text{Re}(L^G) & \text{Re}(L^J) \text{Im}(L^G)\\
          \text{Im}(L^J) \text{Re}(L^G) & \text{Im}(L^J) \text{Im}(L^G)
        \end{bmatrix}
                                          \bigg\} \begin{bmatrix} \text{Re}(f) \\ \text{Im}(f)\end{bmatrix}\\
   = \begin{bmatrix} \text{Re}(S) \\ \text{Im}(S)\end{bmatrix},
\end{multline}
where the multiplication by $(1-\breve y)$ and differentiation $\partial_{\breve y}$ are
understood to represent linear operators on the Legendre Gauss-Lobatto nodal
representation.
We then solve \eqref{eq:H_eqn_linear_solve} by numerically computing the linear
operator along each radial ray and performing an aggregated linear solve via LAPACK.
Boundary conditions are imposed
as usual by setting the
first row of the operands $\text{Re}(S)$ and $\text{Im}(S)$ to the desired
boundary value before the operation, and adjusting the first and $(n+1)$ row of
the linear operator to be equivalent to the first and $(n+1)$ row of the
identity matrix.

\section{Parallelization and modularity} \label{sec:parallelization_modularity}

Because of
the dependence of the gauge transformation at the inner boundary on the
field values at $\mathcal I^+$ needed to establish an asymptotically flat
gauge, the opportunities for subdividing the CCE domain for parallelization
purposes are limited.
However, we are able to take advantage of the
task-based parallelism in SpECTRE to: a) parallelize independent
portions of the CCE information flow, and b) efficiently parallelize the CCE
calculation with a simultaneously running Cauchy simulation.

\subsection{Component construction}
\begin{figure*}[t]
  \centering
  \includegraphics[width=.8\textwidth]{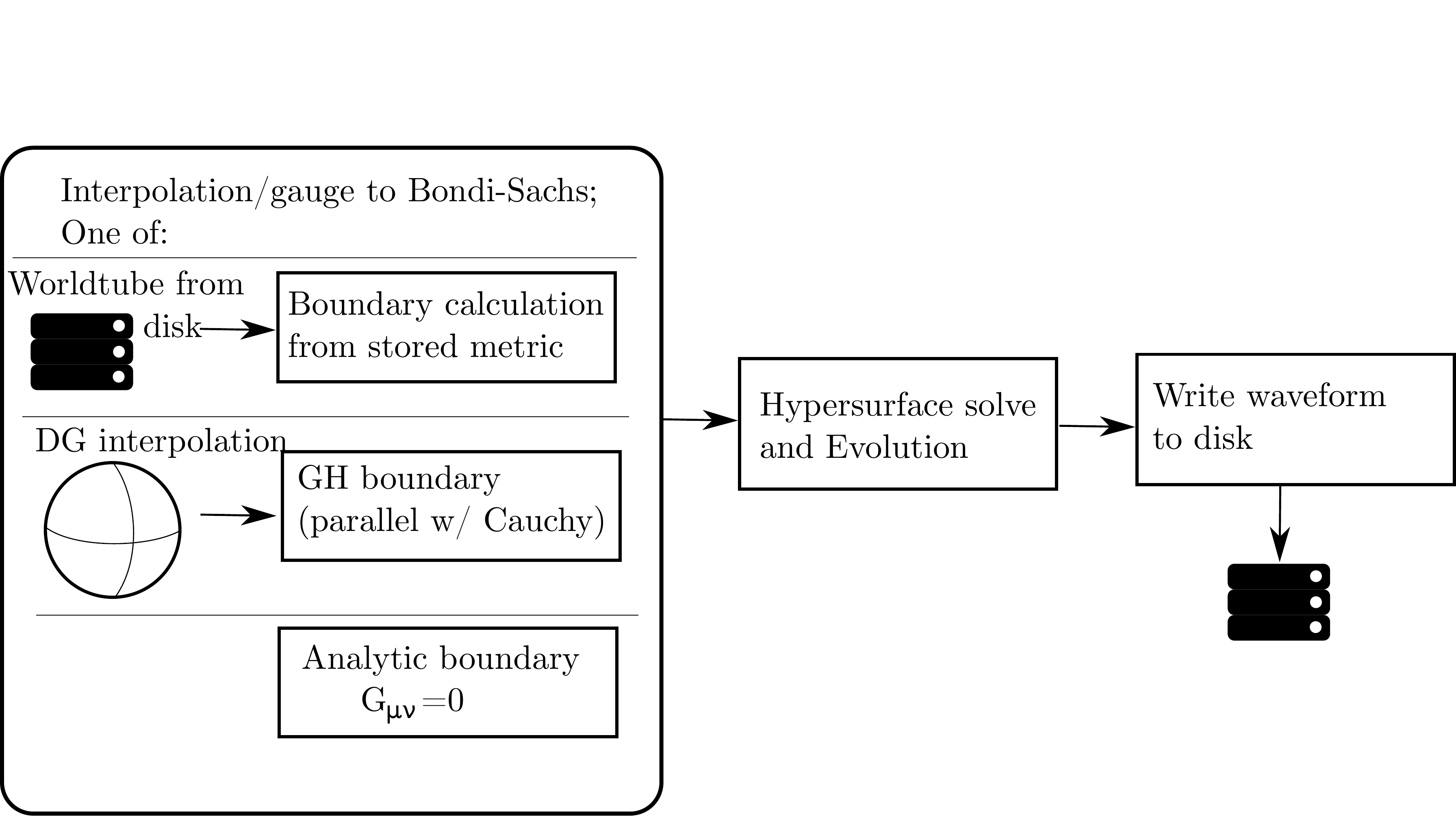}
  \caption{Components of the CCE task-based parallelism system. The worldtube
component (left) is modular and can be switched out according to the desired
source of worldtube data.
We currently support reading worldtube data from disk,
interpolating worldtube data from a simultaneously running Generalized Harmonic
system in SpECTRE, or computing analytic boundary data from a known solution or
approximation to the Einstein field equations.}
\label{fig:components}
\end{figure*}

In SpECTRE, we refer to the separate units of the simulation that may be
executed in parallel via task-based parallelism
as \emph{components}.  For instance, in the near-field
region in which the domain can be parallelized among several subregions of the
domain, each portion of the domain is associated with a component.

For SpECTRE CCE, we use three components (in addition to components that are
used for the Cauchy evolution): one component for the characteristic evolution,
another component dedicated to providing boundary data on the worldtube, and a
third component for writing results to disk.

Much of the efficiency and precision of the SpECTRE CCE system comes from the
ability to cover the entire asymptotic domain from the worldtube $\Gamma$ to
$\mathcal I^+$ with a single spectral domain. In principle, there may be
opportunity to parallelize multiple radial shells of the computation, but in
practice our initial assessments indicated that there would be little gain for
the typical gravitational wave extraction scenario.
First, there is a significant constraint that comes from the asymptotic flatness
condition --- the gauge transformation throughout the domain on a given
hypersurface depends on the asymptotic value $\mathcal U|_{\mathcal I^+}$ on the
same hypersurface, which forces a significant portion of the computation to
serial execution.
Additionally, we have seen very rapid convergence in the number of radial points
used for the CCE system, so it is unlikely that subdividing the domain radially
would offer much additional gain for the typical use case.

Therefore, the entire characteristic
evolution system is assigned to a single component, and
represents the computational core of the algorithm. The evolution component is
responsible for
\begin{itemize}
  \item The angular gauge transformation and interpolation (via Clenshaw recurrence)
  \item The calculation of the right-hand sides of the set of hierarchical
equations \eqref{eq:hierarchy}
  \item The integration of each of the radial ODEs
  \item The time interpolation and preparation of waveform data
\end{itemize}  
The core evolution component performs no reads from or writes to the filesystem,
which ensures that the expensive part of the computation will not waste time
waiting for potentially slow disk operations.

The second component used in CCE is the worldtube component.
A worldtube component is responsible for:
\begin{itemize}
\item Collecting the Cauchy worldtube metric and its derivatives from an assigned data source
\item Interpolating the data to time steps appropriate to the CCE evolution system
\item Performing the transformation to the Bondi-Sachs-like coordinate system on
  the worldtube
\end{itemize}
The user has a choice of several different worldtube components, each of which corresponds to a different source of the metric quantities on the worldtube. Worldtube components are available that:
\begin{itemize}
\item Read worldtube data directly from disk
\item Accept interpolated data from a simultaneously running Cauchy 
  execution in SpECTRE
\item Calculate worldtube data from an analytically determined metric on the boundary
\end{itemize}
Our methods for reading from disk are currently optimized for easily reading
worldtube data written by SpEC, but our worldtube module should accept data from
any code that can produce the spacetime metric and its first derivatives
decomposed into spherical harmonic modes.

Finally, there is a generic observer component that handles the output of the
waveform data to disk.  When CCE is simultaneously running with a Cauchy
evolution, there will be additional components running in parallel with the CCE
components, such as components that perform the Cauchy evolution, components that
search for apparent horizons, and components that write simulation data to disk.
The division of the CCE pipeline into parallel components is illustrated in
Fig.~\ref{fig:components}.

\subsection{Independently stepped interface with Cauchy
simulation} \label{sec:independent_step_size}

Because the Cauchy-characteristic evolution system does not have much
opportunity to parallelize internally, we need to ensure that its serial
execution is optimized.
Our goal is that when running simultaneously with the highly parallel
discontinuous Galerkin system used for the Generalized Harmonic evolution, the
CCE system does not impose any significant runtime penalty.

An important contribution to the efficiency of the CCE system is that the
solutions to the Einstein field equations are smooth and slowly varying in time.  As
a result, the spectral methods used in CCE converge rapidly, and the scales that
we seek to resolve with the time-stepper are primarily on orbital timescales.  
Therefore, we anticipate that the CCE system should be able to take far larger
timesteps than the Generalized Harmonic system running in concert, and it will
be important for the overall efficiency of the extraction pipeline to adjust the
time steps of the CCE evolution independently of the time step of the
Generalized Harmonic system \cite{Lindblom:2005qh}.

Our implementation permits the CCE step size to vary independently of other
  time scales in the simulation, and the step size can be
chosen according to estimates of the time stepper residuals. Those
estimates are frequently obtained by comparing the results of time steppers of
different orders, either via embedded methods \cite{Harier:1993} for substep
integrators, or by varying the number of points used in the arbitrary-sized
multistep methods (LMM) \cite{Throwe:2020} often used in SpECTRE.

The CCE worldtube component receives the metric and its first time derivatives
from the Generalized Harmonic system. It then uses dense output to generate
evenly spaced data sets and barycentric rational interpolation \cite{Floater:2007}
to generate values at the time points required by the CCE evolution
system\footnote{Numerically, only one of either
dense output or barycentric interpolation 
should be sufficient, but we must use both in sequence to satisfy the
constraints of the SpECTRE local time-stepping infrastructure and communication
scheduling.}. This technique
ensures that the interpolated time points will have a precision associated with
the scale of stepper residuals of the Generalized Harmonic system.

\begin{figure}[t]
  \includegraphics[width=0.48\textwidth]{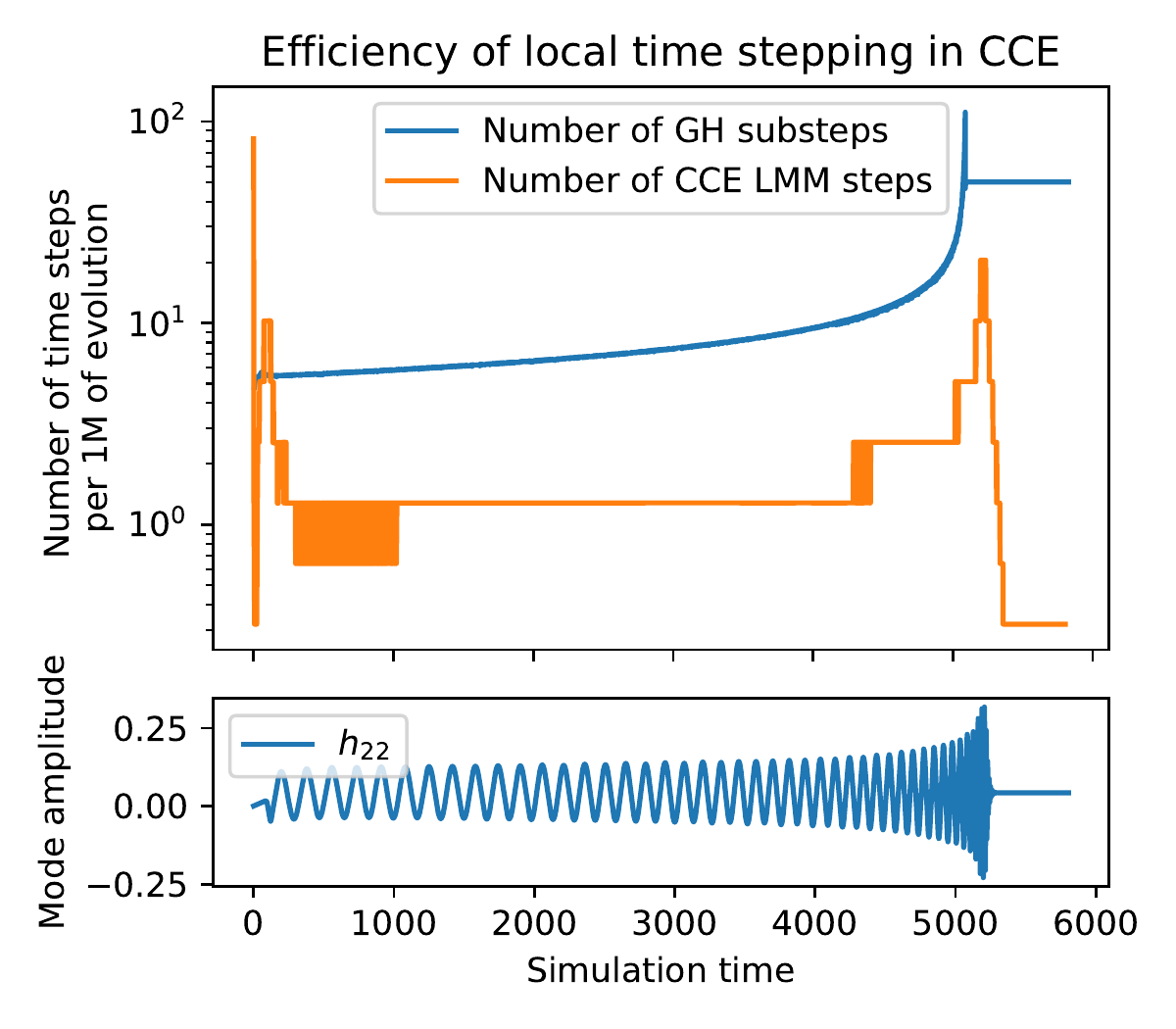}

  \caption{A comparison of the number of substeps taken per $1\,M$ of simulation
time in the SpEC Generalized Harmonic evolution (with Dormand-Prince-5 substep
method) to the number of steps taken in the CCE system (with Adams-Bashforth
order 3 linear multistep method).
We choose to compare the number of substeps to the number of LMM 
steps, as it
most closely represents the factor of speedup in the CCE system as compared to a
system in which CCE is forced to take identical steps to the GH system that
supplies its worldtube data, i.e. a global time-stepping method across all
systems.    
The SpECTRE system chooses steps in discrete factors of 2 as a simplification to
the time-stepping infrastructure and to maintain globally agreed-upon `slabs' of
the evolution, so the time steps chosen in our CCE implementation jump by
factors of 2 during transitions.
The bottom plot shows the dominant gravitational waveform mode for reference.} \label{fig:timesteps}
\end{figure}

To demonstrate the usefulness of our variable step size implementation, we have
performed a simple evolution in SpECTRE using input from a SpEC binary black
hole simulation, and compared the size of the time steps between the SpEC
evolution system and the SpECTRE CCE system.
In Fig.~\ref{fig:timesteps} we show the respective step size of a
globally stepped Generalized Harmonic system in SpEC and the step size of the
SpECTRE CCE system using an adaptive step size based on time stepper residuals.

For the evolution system in SpECTRE, we will have the opportunity to perform
local time-stepping for separate elements in the Generalized Harmonic domain as
well, which will allow the elements in direct communication with the CCE system
to take larger steps.  However, even for modest resolution in the SpECTRE
Generalized Harmonic system, and for a 100M worldtube radius
, we should expect
the Generalized Harmonic system to still take $\mathcal{O}(10)$ steps per $M$ of
evolution for an Adams-Bashforth order 3 scheme, so the CCE system should still
benefit from an independently chosen step size.

The examples in this section emphasize the value in permitting the CCE system to
choose its own step size ---
the smoothness of the solution across the asymptotic
null hypersurface ensures that CCE can comfortably take far larger time steps
than its partner Generalized Harmonic system.
The larger time steps then permit either a far faster extraction in the case of
a standalone CCE run, or permit the CCE system to make negligible impact on the
overall runtime when evolved in tandem with the Generalized Harmonic evolution
in SpECTRE.

\section{Battery of tests} \label{sec:battery_of_tests}

\subsection{Barkett test collection}

In \cite{Barkett:2019uae}, we described a series of demanding tests for verifying
the correctness and efficiency of the SpEC implementation of CCE.
We have reimplemented all five of those analytic tests for SpECTRE CCE, and
similarly verified the correctness and convergence properties of the new
implementation.
Here, we briefly summarize the salient features of each of the test cases and
present the results from applying the collection of tests to our new
implementation in SpECTRE.
Please refer to \cite{Barkett:2019uae} for complete details regarding the
formulation of the tests.

Each test generates Cauchy worldtube metric and its
derivatives on a chosen worldtube and uses a custom version of the CCE
worldtube component to provide the Cauchy worldtube
metric and its derivatives to the characteristic evolution component.  Otherwise, the
remainder of the pipeline operates precisely as it would if extracting waves
from data generated by a full Cauchy evolution (see Fig.~\ref{fig:components}).
The analytic tests provide a prediction for the asymptotic Bondi-Sachs news
function, which is then compared against the extracted news function to
determine a residual and evaluate the precision of the CCE system.

\begin{figure}[t]
  \includegraphics[width=.48\textwidth]{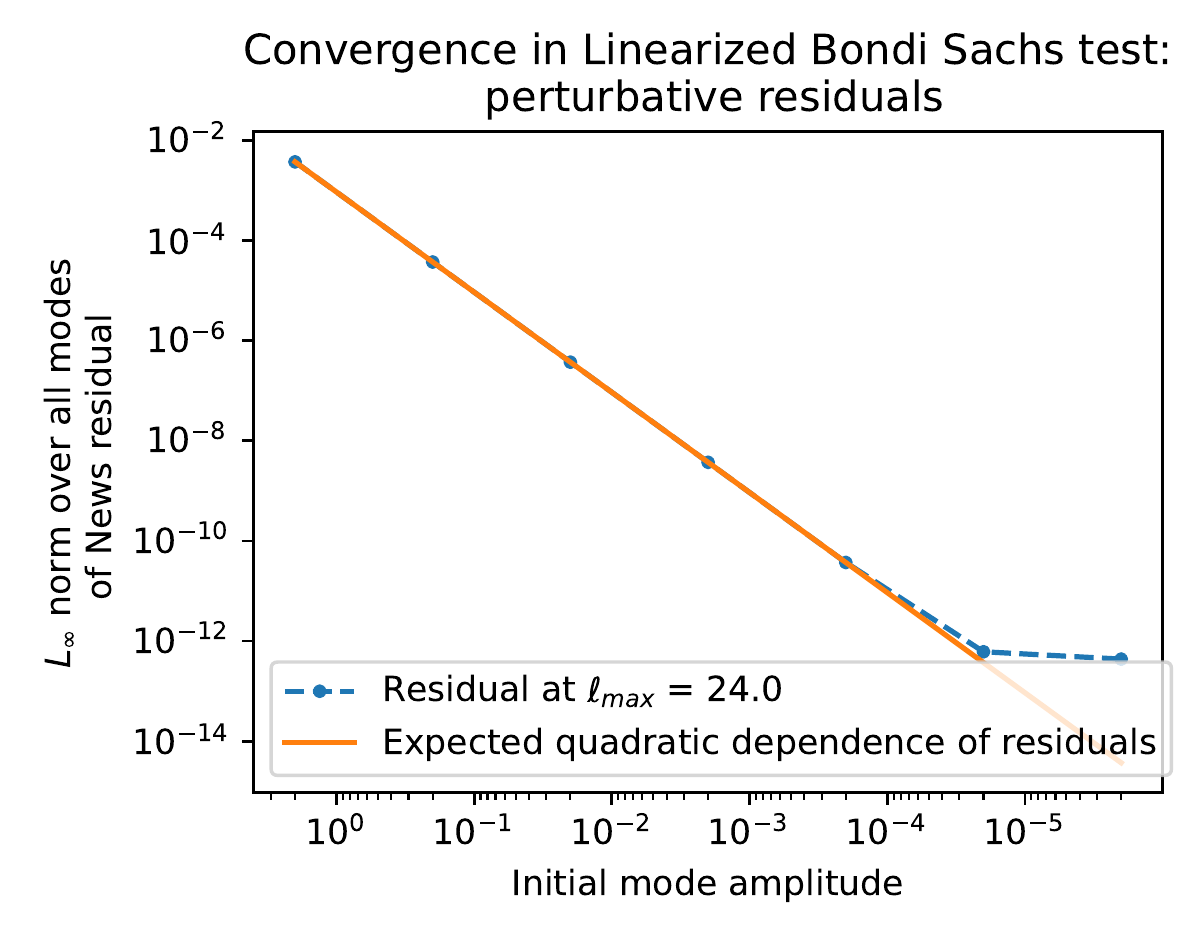}
  \caption{Residual obtained by subtracting the numerical SpECTRE CCE news from
the linearized Bondi-Sachs news.
The residual follows closely the expected perturbative residual $\propto
\alpha^2$ for amplitude $\alpha$, until the time stepper residual dominates
at $\sim 10^{-12}$ (The absolute tolerance of the time stepper
is chosen to be $10^{-13}$ in these tests
and run for a duration of $5/\nu$).
} \label{fig:linearized_bondi_sachs_convergence}
\end{figure}

\begin{figure}[t]
  \includegraphics[width=.48\textwidth]{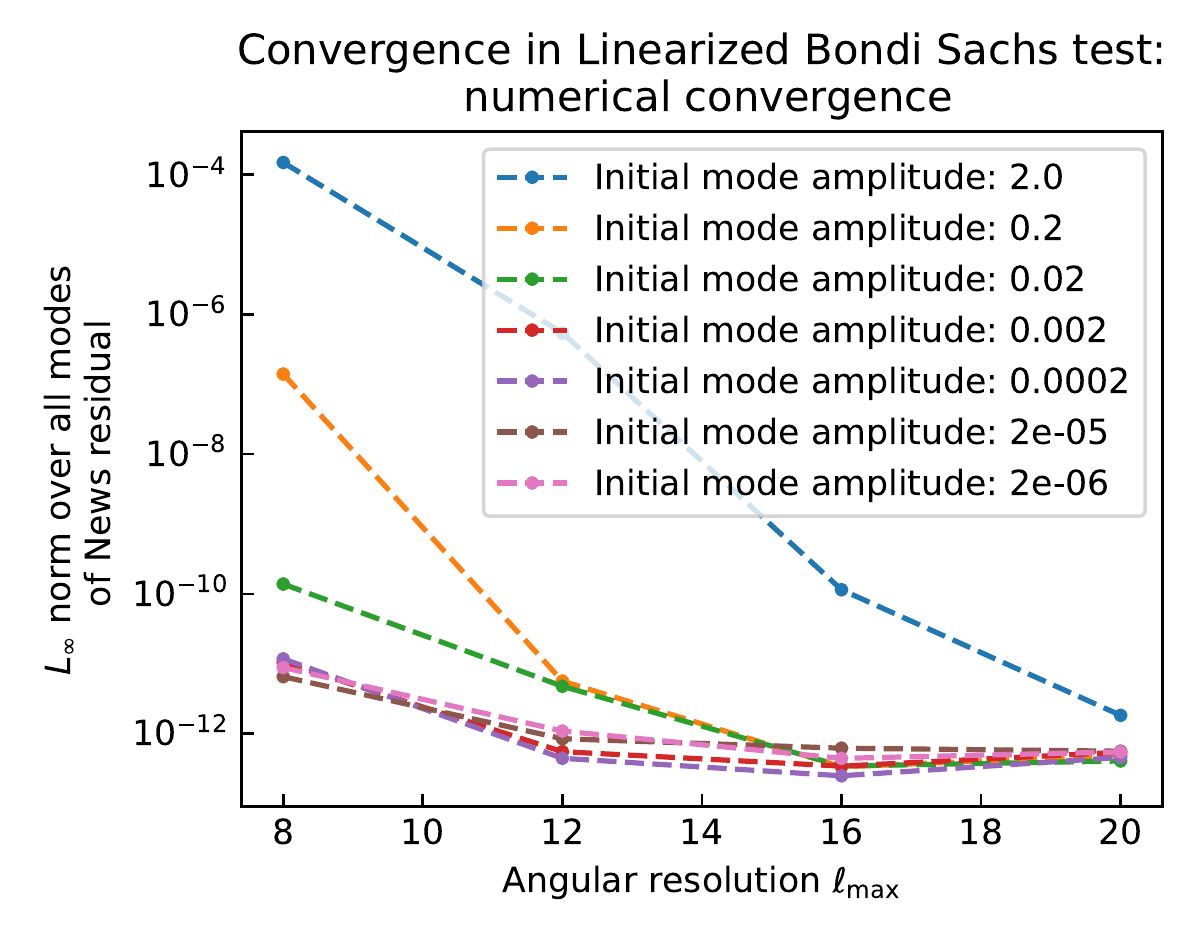}
  \caption{Numerical residual in the Linearized Bondi-Sachs test, obtained by
subtracting the extracted news from its value at the maximum resolution
($\ell_{\text{max}}=24$) for each given amplitude.
  } \label{fig:linearized_bondi_sachs_numerical_residuals}
\end{figure}

\paragraph{Linearized Bondi-Sachs:} This solution expands the
Bondi-Sachs metric \eqref{eq:bs_metric} around the flat space solution, and was first
derived in \cite{Bishop:2004ug}. The spin-weighted scalars that determine the metric are
expanded in modes as
\begin{subequations}
\begin{align}
  J_{\text{lin}\,\ell m} &= \sqrt{(\ell + 2)!/(\ell - 2)!}\; {}_2 Z_{\ell m} \text{Re}[J_{\ell}(r) e^{i \nu u}]\\
  U_{\text{lin}\,\ell m} &= \sqrt{\ell (\ell + 1)}\; {}_1 Z_{\ell m} \text{Re}[U_{\ell}(r) e^{i \nu u}]\\
  \beta_{\text{lin}\,\ell m} &= {}_0 Z_{lm} \text{Re}[\beta_{\ell}(r) e^{i \nu u}]\\
  W_{\text{lin}\,\ell m} &= {}_0 Z_{lm} \text{Re}[W_{\ell}(r) e^{i \nu u}],
\end{align}
\end{subequations}
where $\nu$ is a user-defined frequency and each radially dependent $\ell$-mode of
the solution is specified by analytic
calculation via the expansion of the Einstein field equations in the Bondi-Sachs
gauge, and the spin-weighted spherical harmonic functions ${}_s Z_{\ell m}$ from
\cite{Bishop:2004ug} are
\begin{align}
  {}_s Z_{\ell m} = \begin{cases}
    \frac{i}{\sqrt{2}} \left[(-1)^m{}_s Y_{\ell m} -  {}_s Y_{\ell -m} \right],& m < 0, \\
    {}_s Y_{\ell 0}, & m = 0, \\
    \frac{1}{\sqrt{2}} \left[{}_s Y_{\ell m} + (-1)^m {}_s Y_{\ell -m} \right],& m > 0. \end{cases}
\end{align}
The asymptotic news function is then
\begin{align}
  N_{\text{lin}\, \ell m} =& \text{Re}\left[e^{i \nu u} \lim_{r\rightarrow\infty}\left(\frac{\ell(\ell + 1}{4} J_\ell - \frac{i \nu r^2}{2} J_{\ell, r} + \beta_\ell\right)\right]\notag\\
  &\hspace{5mm} \times\sqrt{\frac{(\ell + 2)!}{(\ell - 2)!}} {}_2 Z_{\ell m}.
\end{align}

As in the SpEC implementation, we consider only $\ell=2$ and $\ell=3$ modes, for
which the full radial dependence is given in \cite{Bishop:2004ug,Barkett:2019uae}. Because
the above linearized Bondi-Sachs expressions are valid only to first order in
the perturbation amplitude $\alpha$, but CCE evolves the full nonlinear Einstein
equations, the difference between the linearized solution and CCE should depend
quadratically on the amplitude $\alpha$.
  In
  Fig.~\ref{fig:linearized_bondi_sachs_convergence}, we plot this difference
    versus $\alpha$, and we recover the expected quadratic
    dependence, so that  the relative residual is
proportional to the perturbation amplitude $\alpha$. Figure
\ref{fig:linearized_bondi_sachs_numerical_residuals} shows the convergence of
the CCE news with angular resolution $\ell_{\text max}$: plotted is the
difference between the CCE news at a given $\ell_{\text max}$ and the CCE news
at $\ell_{\text max} = 24$.

\begin{figure}[t]
  \includegraphics[width=.48\textwidth]{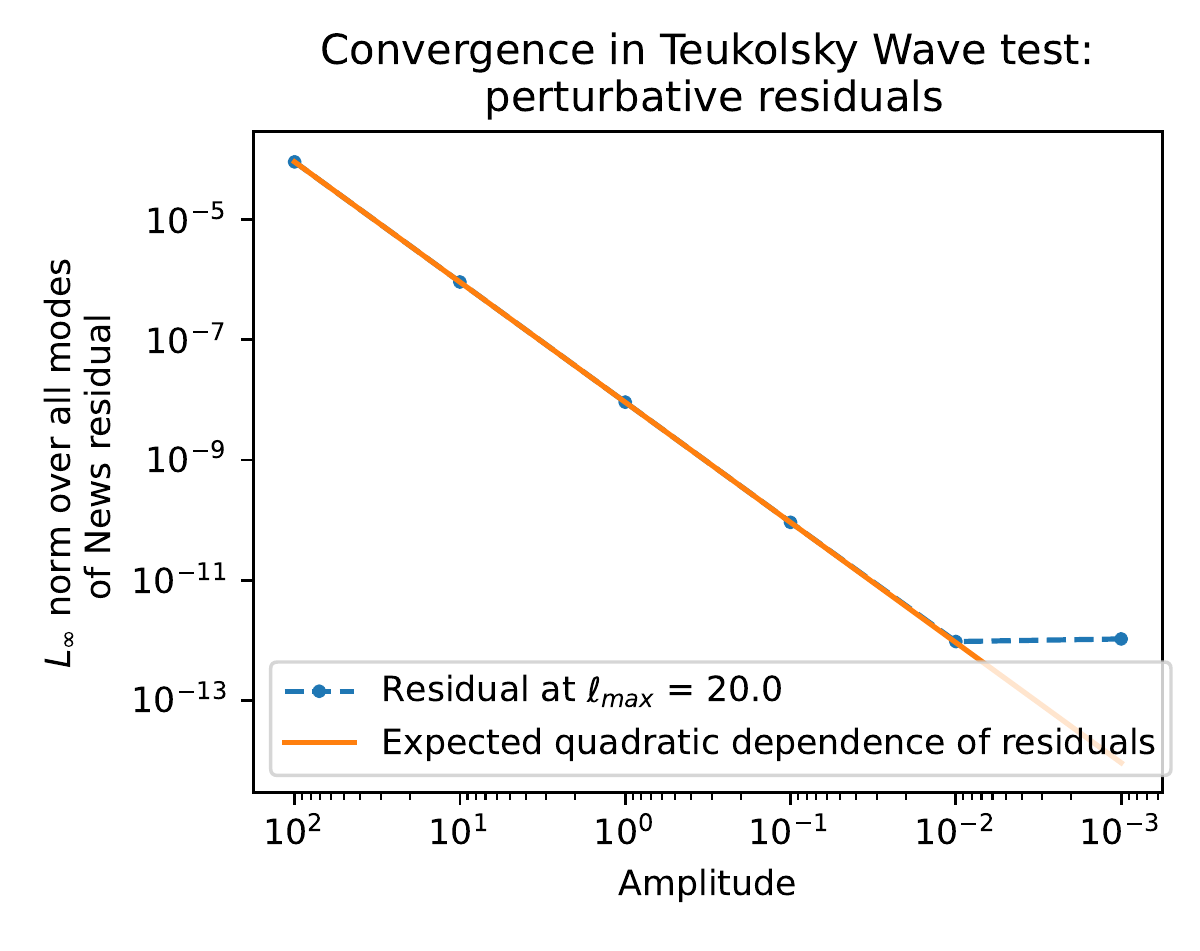}
  \caption{Residual obtained by subtracting the SpECTRE
    CCE news from 
      the news computed from an
    $(\ell, m) = (2, 0)$ Teukolsky
    wave.
    The residual follows closely the expected perturbative residual $\propto
    \alpha^2$ for amplitude $\alpha$, until the time stepper residual dominates
at $\sim 10^{-12}$ (The absolute tolerance of the time stepper is chosen to be $10^{-13}$ in these tests
and run for duration $5 \tau$).
  } \label{fig:teukolsky_convergence}
\end{figure}

\begin{figure}[t]
  \includegraphics[width=.48\textwidth]{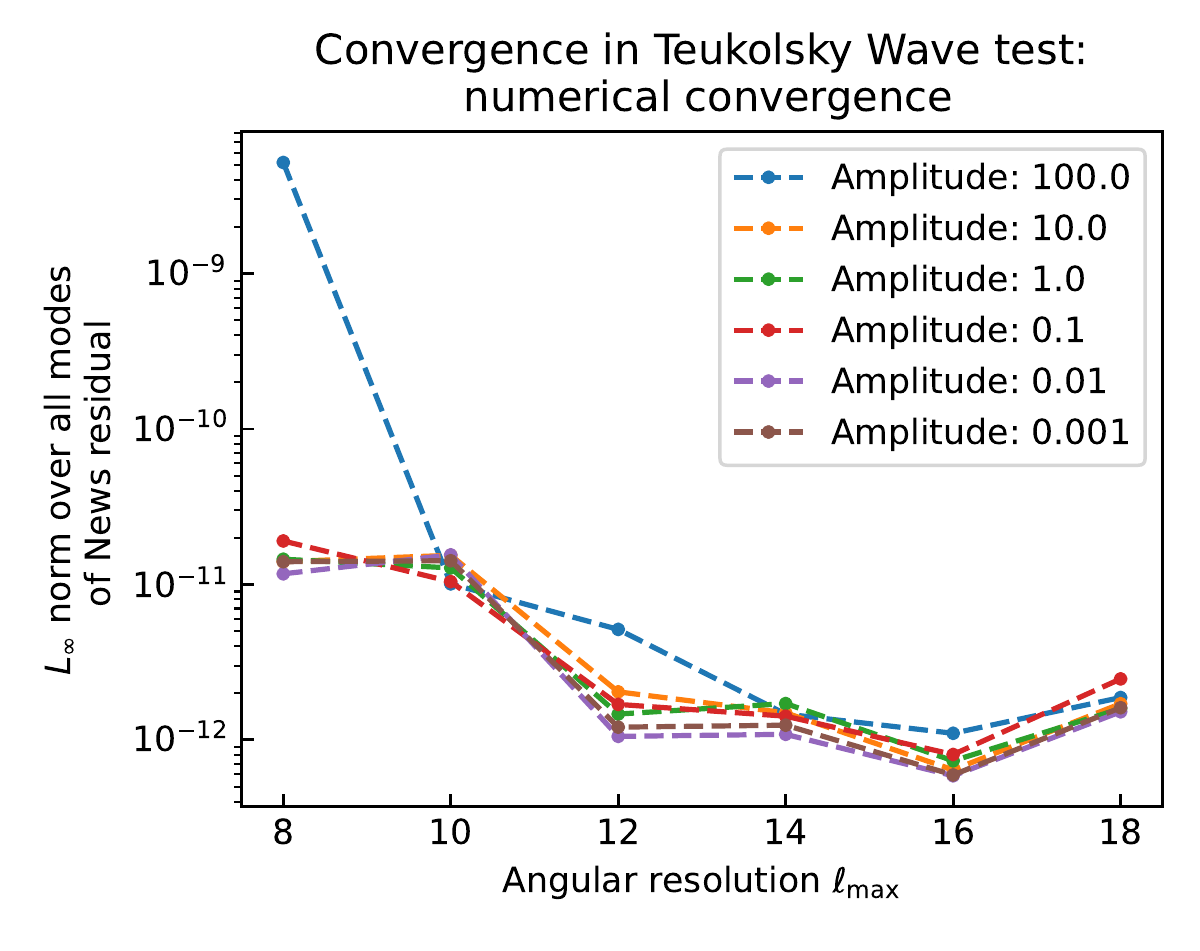}
  \caption{Numerical residual in the Teukolsky wave test, obtained by
    subtracting the extracted news from its value at the maximum resolution
    ($\ell_{\text{max}}=20$) for each given amplitude.} \label{fig:teukolsky_numerical_residuals}
\end{figure}

\paragraph{Teukolsky wave:} A linearized perturbation on a flat background is
evaluated on the worldtube and compared against the predicted asymptotic news.
We use the outgoing form of the linearized metric given in \cite{Teukolsky:1982nz, Barkett:2019uae}:
\begin{align}
  ds^2 =& -dt^2 + (1 + f_{rr}) dr^2 + 2 B f_{r \theta} r dr d\theta \notag\\
        &+ 2Bf_{r \phi} r \sin \theta dr d \phi + \left(1 + C f_{\theta \theta}^{(1)}
          + A f_{\theta \theta}^{(2)}\right) r^2 d\theta^2 \notag\\
        &+ 2 (A - 2 C) f_{\theta \phi} r^2 \sin \theta d \theta d \phi \notag\\
        &+ \left(1 + C f_{\phi \phi}^{(1)} + A f_{\phi \phi}^{(2)}\right) r^2 \sin^2 \theta d\phi^2,
\end{align}
where the functions $A$, $B$, and $C$ are determined by the arbitrary wave
profile function $F(u) = F(t - r)$:
\begin{subequations}
\begin{align}
  A &= \frac{1}{r^3} \left(\partial_u^2 F + 3 r^{-1} \partial_u F + 3 r^{-2} \partial_u F\right) \\
  B &= -\frac{1}{r^2} \left(\partial_u^3 F + 3 r^{-1} \partial^2_u F + 6 r^{-2} \partial_u F + 6 r^{-3} \partial_u F\right) \\
  C &= \frac{1}{4 r} \big(\partial_u^4 F + 2 r^{-1} \partial^3_u F + 9 r^{-2} \partial^2_u F\notag\\
    &\hspace{2cm} + 21 r^{-3} \partial_u F + 21 r^{-4} F\big),
\end{align}
\end{subequations}
and the $f_{i j}^{(n)}$ functions are tensor harmonic functions determined by
the choice of ${}_s Y_{\ell m}$ modes.
We follow \cite{Barkett:2019uae} and choose a strictly outgoing ${}_2 Y_{2 0}$
mode, and for that choice of solution, the asymptotic news is
\begin{equation}\label{eq:TeukolskyWaveNews}
  N = \frac{3}{4} \sin^2\theta \,  \partial_u^5 F(u).
\end{equation}
We also choose a Gaussian wave profile $F(u) = \alpha e^{-u^2/\tau^2}$ with
amplitude $\alpha$ and width $\tau$.

As in the Linearized Bondi-Sachs solution, the analytic solution for the
Teukolsky wave generates a nontrivial waveform, but the solution is
perturbative. Because CCE evolves the full nonperturbative Einstein equations,
the difference between CCE and the perturbative solution should scale as
$\alpha^2$.
We show the convergence of the residual with diminishing amplitude in Fig.~\ref{fig:teukolsky_convergence},
and in Fig.~\ref{fig:teukolsky_numerical_residuals} we show the convergence of
the numerical residuals determined by comparing to the highest resolution run
conducted ($\ell_{\text{max}} = 20$).

\def\arraystretch{1.4}
\begin{table*}
  \begin{tabular}{@{\hspace{1em}}c@{\hspace{1em}}@{\hspace{1em}}c@{\hspace{1em}}@{\hspace{1em}}c@{\hspace{1em}}}
    Test name & Parameters tested & Maximum residual \\
    \hline
    Rotating Schwarzschild & $\ell_{\text{max}} \in [8, 24]; \omega \in [0.1, 0.8]$ & $2.94\times 10^{-11}$ \\
    Gauge Wave &$\ell_{\text{max}} \in [8,24]; \alpha \in [0.01, 10.0]$ & $4.05\times 10^{-12}$\\
    \hline
    \hline
\end{tabular}
\caption{Maximum residuals across the explored parameter space for the rapidly
  converging test cases.
}
\label{tab:quickly_converging_test_cases}
\end{table*}

\paragraph{Rotating Schwarzschild:} We generate worldtube data from the
Schwarzschild metric in Eddington-Finkelstein coordinates, with an angular
coordinate transformation $\phi \rightarrow \phi + \omega u$ applied:
\begin{align}
  ds^2 =& -\left(1 - \frac{2 M}{r} - \omega^2 r^2 \sin^2 \theta\right) du^2 - 2 du dr \notag\\
  &+2 \omega r^2 \sin^2 \theta du d\phi + r^2 \sin^2 \theta d\Omega^2.
\end{align}
This test case produces no waves, so we expect to recover news $N = 0$ from the CCE
system. For this test case, we find extremely small residuals across the
parameter space that we explored. No run resulted in any mode of the news
exceeding  ${}\sim 10^{-11}$;
this test case is included in summary
table~\ref{tab:quickly_converging_test_cases}.

\begin{figure}[t]
  \includegraphics[width=.48\textwidth]{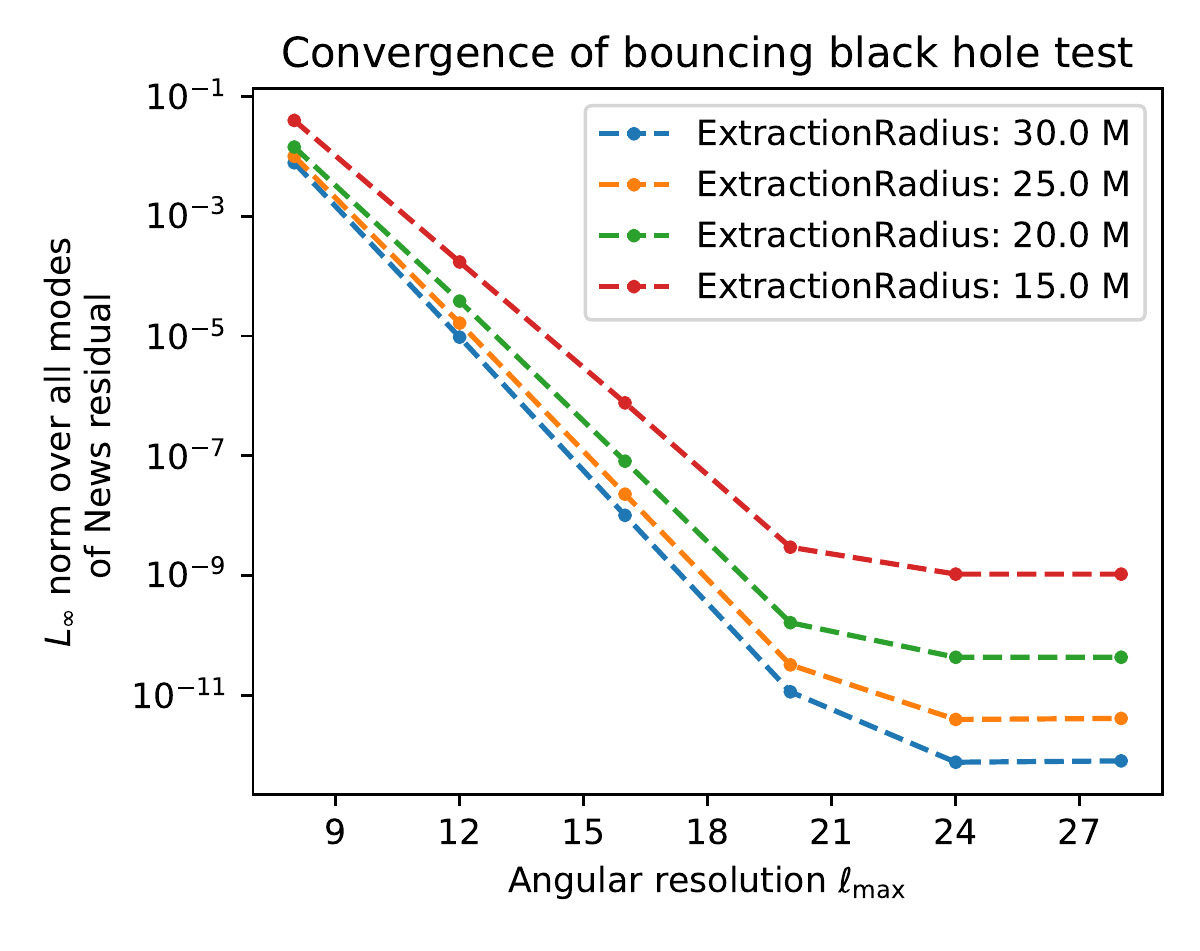}
  \caption{Convergence of SpECTRE CCE for the bouncing black hole test across
    various extraction radii.
    Here, we have set the absolute tolerance of the stepper residual to
    $10^{-12}$ to ensure that the residuals are associated only with the spatial
    resolution. At an extraction radius of $15 M$,
    the convergence saturates
    slightly early, at around $\sim 10^{-9}$. However, at even modestly larger
    extraction radii, the SpECTRE CCE system approaches the expected truncation
    scale of the spectral scheme.}
  \label{fig:spectre_bouncing_bh_convergence}
\end{figure}
\paragraph{Bouncing Schwarzschild black hole:} The worldtube data for the
bouncing black hole test is similar conceptually to the Rotating Schwarzschild
test.
However, instead of performing an angular coordinate transformation, here we
apply a time dependent linear transformation to the Kerr-Schild coordinate
system $(t, x, y, z)$:
\begin{equation}
  x\rightarrow x + a \sin^4(2 \pi t / b).
\end{equation}
As in the Rotating Schwarzschild test, the expected result of the CCE system is
zero news, since the solution is simply Schwarzschild in an oscillating coordinate system.
For our tests, we choose an oscillation amplitude $a = 2M$ and period $b = 40M$.

In practice, the bouncing black hole test has proven to be a far more demanding
test of the CCE wave extraction system than many of the other tests.
A naive examination of individual Bondi-Sachs scalars in this scenario would
lead one to believe that there is wave content in the system --- it is only
through the full nonlinear simulation that the CCE system is able to resolve the
motion as a pure-gauge effect and cancel all of the contributions in the final
value of the news.
We show the convergence of the SpECTRE CCE system for the bouncing Schwarzschild
black hole test in Fig.~\ref{fig:spectre_bouncing_bh_convergence}.

\begin{figure}[t!]
  \includegraphics[width=.48\textwidth]{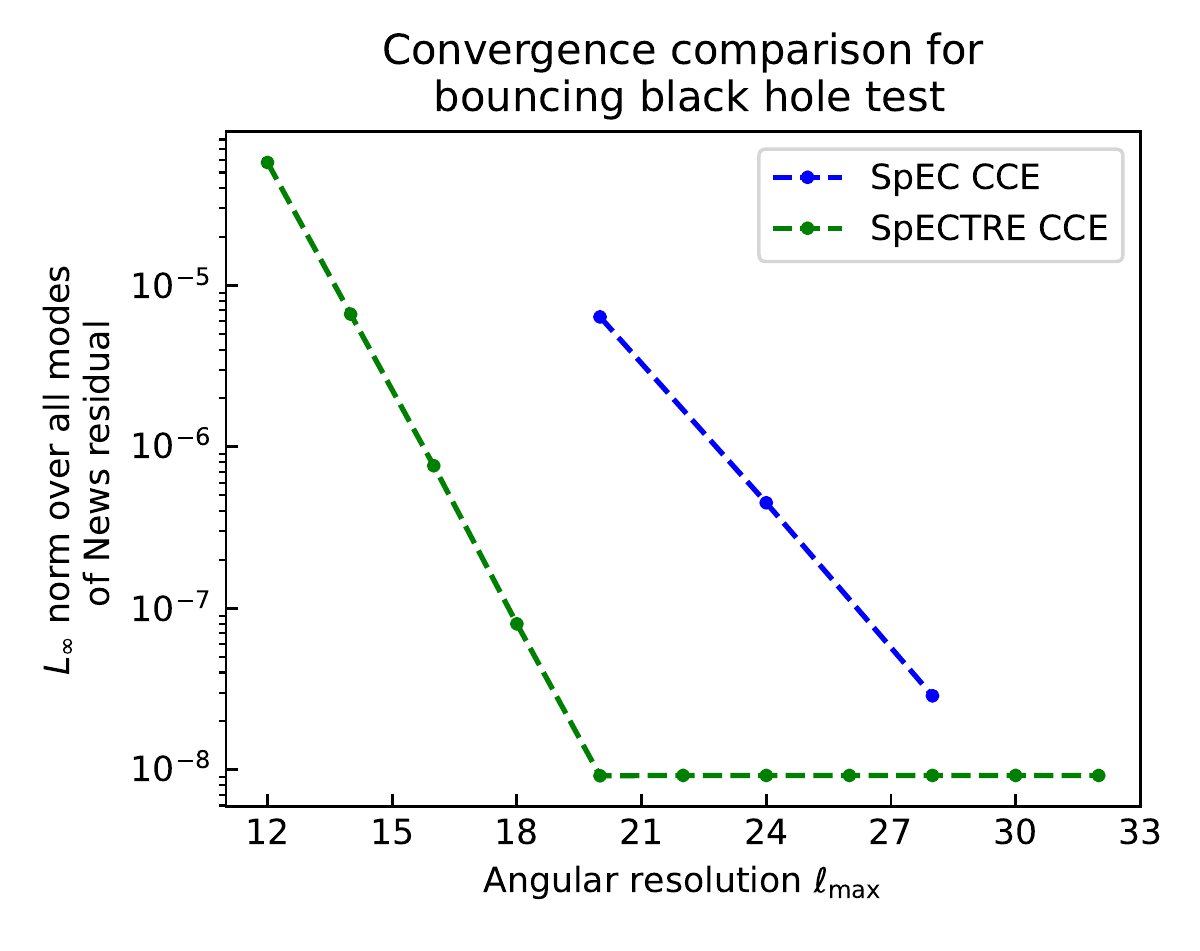}
  \caption{Convergence of SpEC and SpECTRE CCE systems when using matched
    parameters for the collection of tests used for the bouncing black hole in
    \cite{Barkett:2019uae}. The residual floor reached at $\ell_{\text{max}}=20$ is dominated by
    the absolute stepper residual. Fig.~\ref{fig:spectre_bouncing_bh_convergence}
    shows convergence over several extraction radii for SpECTRE alone, for runs in
    which we use a more aggressive stepper residual and achieve a finer precision.}
  \label{fig:bouncing_bh_convergence_comparison}
\end{figure}

\begin{figure}[t!]
  \includegraphics[width=.48\textwidth]{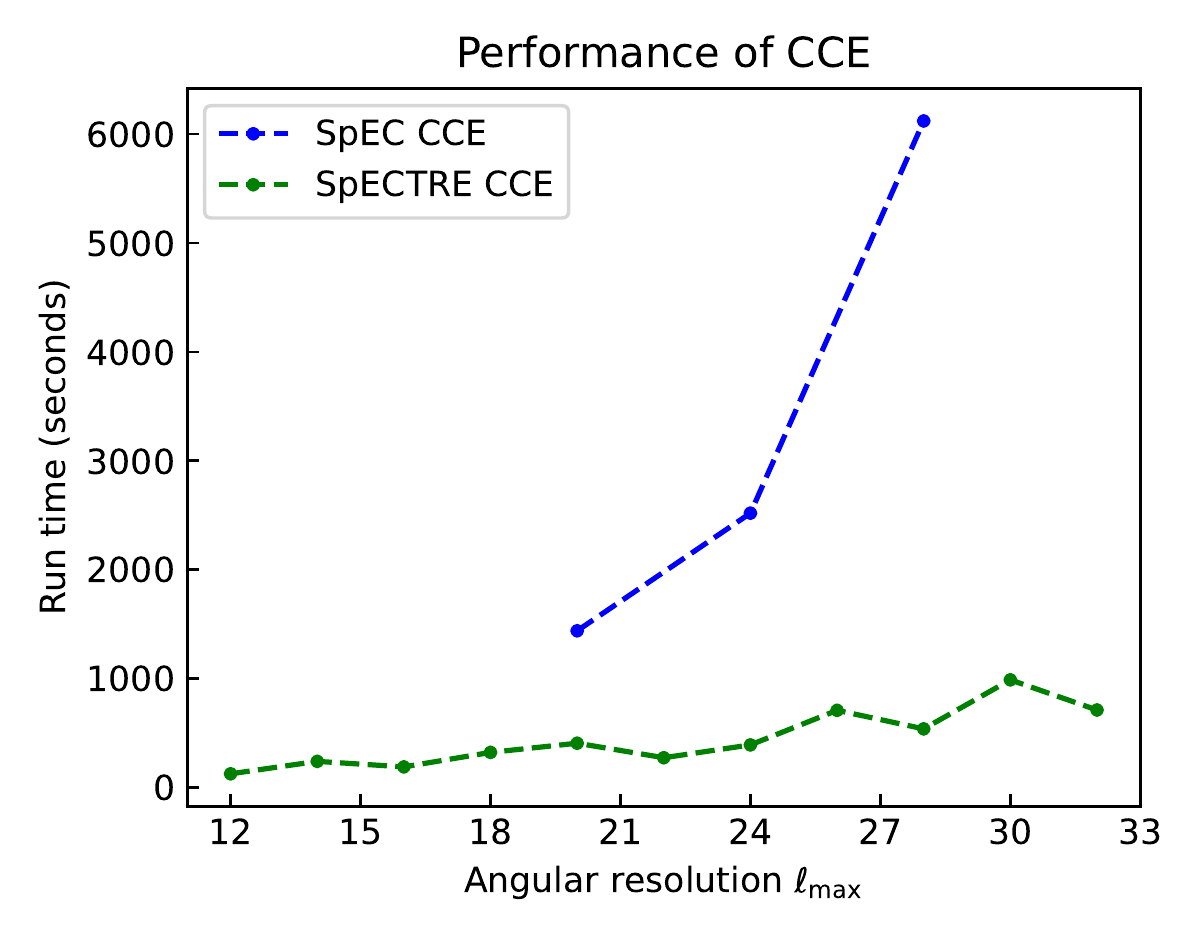}
  \caption{Performance comparison between the SpEC and SpECTRE CCE systems
    applied to the bouncing black hole test.
    We find that the SpECTRE performs considerably better for a comparable
    selection of simulation and stepper parameters.
    The non-monotonicity of the SpECTRE execution time likely arises from the
    dependence of the core SWSH operations performed via {\sc libsharp}, and the
    corresponding dependence on fast Fourier transform algorithms that perform
    better for some mode numbers than others.}
  \label{fig:bouncing_bh_perfomrance_comparison}
\end{figure}

Because the bouncing black hole test has proven to be such a thorough test of
CCE, we have chosen this test case as a benchmark system to compare the SpEC and
SpECTRE simulation codes, both for speed and precision.
In Figs.~\ref{fig:bouncing_bh_convergence_comparison} and
\ref{fig:bouncing_bh_perfomrance_comparison} we show the relative precision and
wallclock execution times for the two systems, for similar parameters of the
test system.

We use a somewhat more demanding error measure than in the previous publication
\cite{Barkett:2019uae}: we take the maximum error over all extracted modes,
instead of examining particular modes and potentially neglecting the highest
modes that can accumulate nontrivial value.
The angular resolution plotted here is the $\ell_{\text{max}}$ used during the
worldtube transformation phase of the computation, as that is the most demanding
part of the calculation for angular resolution. The SpEC implementation
uses twice the angular resolution on the worldtube boundary as on the volume,
and keeps all but the top two modes from the volume when writing to disk. The
SpECTRE implementation uses the same resolution on the boundary as in the
volume, and for these runs we write the same set of modes as SpEC for
consistency in the comparison.
The SpEC runs are the same three runs as were used in the performance and
convergence tests reported in \cite{Barkett:2019uae}.

We find that the SpECTRE implementation enjoys significantly better precision,
executes more quickly, and scales to high resolutions more gracefully than our
previous SpEC implementation.
At the highest resolution that we anticipate will be practical for the typical
binary black hole wave extraction, $\ell_{\text{max}}=24$, we find that our new
SpECTRE implementation performs ${}\sim6\times{}$ faster.

\paragraph{Gauge wave:} The final test in the collection of analytic tests
assembled in \cite{Barkett:2019uae} is an exact wavelike solution that is
equivalent to a gauge transformation applied to the Schwarzschild spacetime.
The metric is constructed by applying the coordinate transformation $v = t + r +
F(t - r)/r$, where the function $F(u)$ is the wave profile function.
Following the coordinate transformation, the Schwarzschild metric is
\begin{widetext}
\begin{align}
  ds^2 =& -\left(1 - \frac{2 M}{r}\right)\left(1 + \frac{\partial_u F}{r}\right)^2 dt^2 + 2\left(1 + \frac{\partial_u F}{r}\right)\bigg[\frac{2 M}{r} + \left(1 - \frac{2 M}{r}\right)\left(\frac{\partial_u F}{r} + \frac{F}{r^2}\right)\bigg] dr dt\notag\\
   &+ \left(1 - \frac{\partial_u F}{r} - \frac{F}{r^2}\right)\left[1 + \frac{2 M}{r} + \left(1 - \frac{2 M}{r}\right)\left(\frac{\partial_u F}{r} + \frac{F}{r^2}\right)\right]
   + r^2 d\Omega^2.
\end{align}
\end{widetext}
For our implementation, as in \cite{Barkett:2019uae}, we use a sine-Gaussian wave
profile $F(u)=A\sin(\omega u) e^{-(u - u_0)^2/\tau^2}$, with frequency $\omega = 0.5$, duration $\tau = 10.0$, and peak time $u_0 = 25.0$.
Once again, we expect to recover zero news, as there is no physical wave content
in the gauge wave spacetime.
The SpECTRE CCE implementation performs well; across the parameter space that we
tested, we find no residuals greater than $4.05\times10^{-12}$. The test results are
summarized in Table~\ref{tab:quickly_converging_test_cases}.

\subsection{Robinson-Trautman solution}

\begin{figure}[t]
  \includegraphics[width=0.48\textwidth]{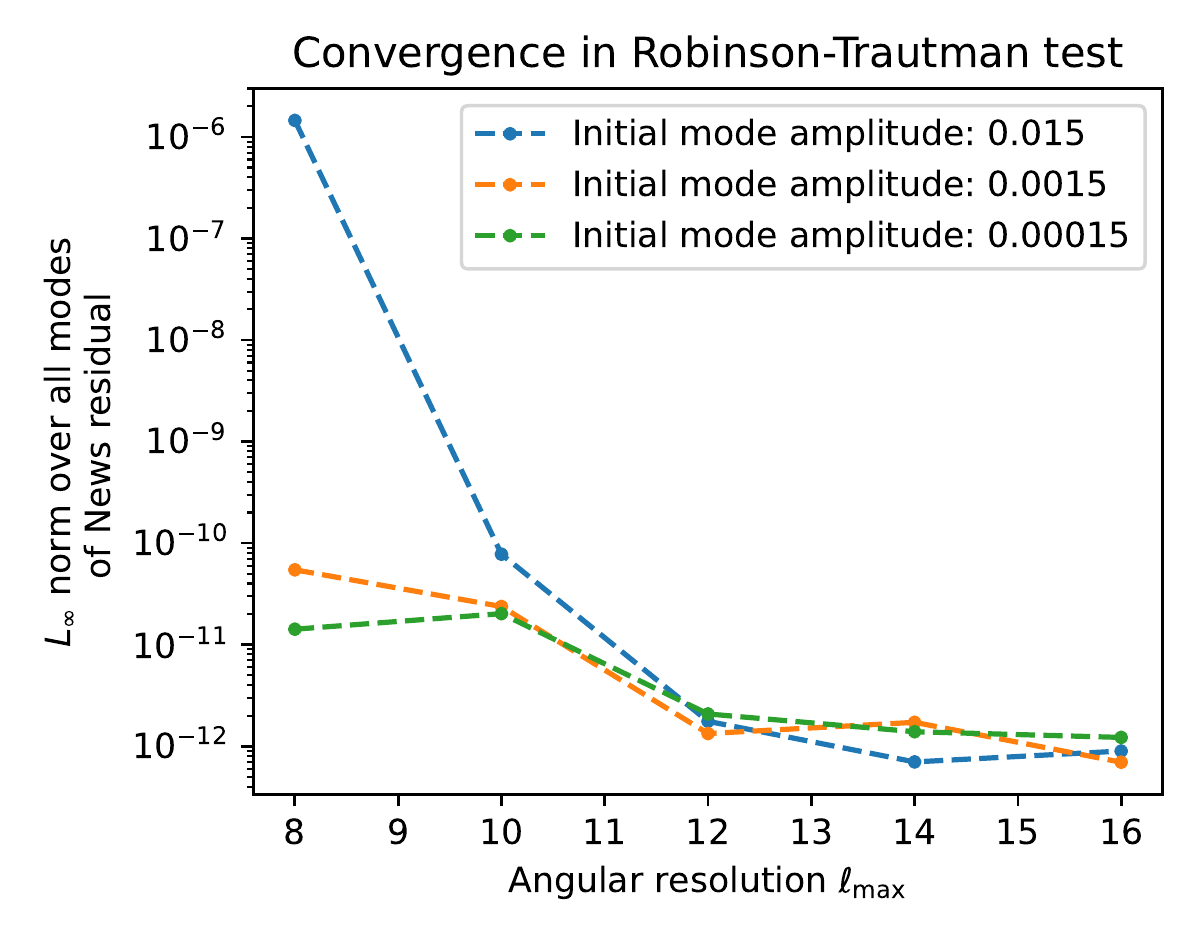}
  \caption{Residuals for the Robinson-Trautman test, computed by subtracting the
    extracted news from the analytic prediction of the news.
  }
  \label{fig:RT_convergence}
\end{figure}

In addition to the five tests constructed previously, we have implemented an
analytic test of SpECTRE CCE based on the Robinson-Trautman metric
\cite{Derry:1969hqa}.
We specialize the construction of the Robinson-Trautman metric to the case in
which there is no coordinate singularity in the asymptotic domain simulated in
CCE.
Unlike the tests in the above collection, the Robinson-Trautman solution both generates
nontrivial asymptotic Bondi-Sachs News, and does not rely on any linearized
approximations.
However, it is not a fully analytic solution---a single scalar variable needs
to be numerically evolved on the worldtube surface to determine the full
Bondi-Sachs metric.

The specialization of the Robinson-Trautman solution that we use depends on the
scalar spin-weight zero surface quantity that we denote $\omega_{\text{RT}}$.
The Robinson-Trautman metric solution takes the form \cite{Derry:1969hqa},
\begin{align} \label{eq:RT_metric}
  ds^2 =& -\left[\left(r W + 1\right) \omega_{\text{RT}} - r^2 U \bar U\right](dt - dr)^2 \notag\\
        &- 2 \omega_{\text{RT}}(dt - dr) dr - 2 r^2 U^A q_{A B} dx^B (dt - dr)\notag\\
        &+ r^2 q_{A B} dx^A dx^B,
\end{align}
where $q_{A B}$ represents the angular unit sphere metric, and the Bondi-Sachs scalars and angular tensors are defined in terms of the Robinson-Trautman scalar $\omega_{\text{RT}}$ as
\begin{subequations}
  \begin{align}
    W &= \frac{1}{r}\left(\omega_{\text{RT}} + \eth \bar \eth \omega_{\text{RT}} - 1 \right) - \frac{2}{r^2 \omega^2_{\text{RT}}},\\
    U &\equiv U^A q_A = \frac{\eth \omega_{\text{RT}}}{r}.
  \end{align}
\end{subequations}

The Robinson-Trautman scalar may be chosen arbitrarily for a single initial retarded
time $u = u_0$, and at all later times, is computed by integrating the evolution equation
\begin{equation}
  \partial_u \omega_{\text{RT}} = - \left[\omega^4_{\text{RT}} \eth^2 \bar \eth^2 \omega_{\text{RT}}
    - \omega^3_{\text{RT}} \left(\eth^2 \omega_{\text{RT}}\right)\left(\bar \eth^2 \omega_{\text{RT}}\right) \right].
\end{equation}
The news for the solution is
\begin{equation}
  N = \frac{\bar \eth \bar \eth \omega_{\text{RT}}}{\omega_{\text{RT}}}.
\end{equation}

We have performed the Robinson-Trautman test over a range of angular resolutions
and for various initial magnitudes of the Robinson-Trautman scalar
$\omega_{\text{RT}}$, and the rapid convergence for this test case is shown in
Fig.~\ref{fig:RT_convergence}.
For our tests, we choose a starting $\omega_{\text{RT}}$ with nonzero modes:
  \begin{align}
    a_{0\, 0} = A \qquad a_{1\, -1} = 4 A / 3 \qquad a_{1\, 0} = A / 3 (2 + i) \notag\\ a_{1\, 1} = 4 A  (1 + 2 i)/3 \qquad a_{2\, -2} = A (5 + 2 i)/3,
  \end{align}
where $A$ is the initial mode amplitude that is varied in Fig.~\ref{fig:RT_convergence}.
All other modes of $\omega_{\text{RT}}$ are zero at $t = 0$.

\section{Binary black hole simulation trials} \label{sec:bbh_trials}

\begin{figure*}
  \includegraphics[width=\textwidth]{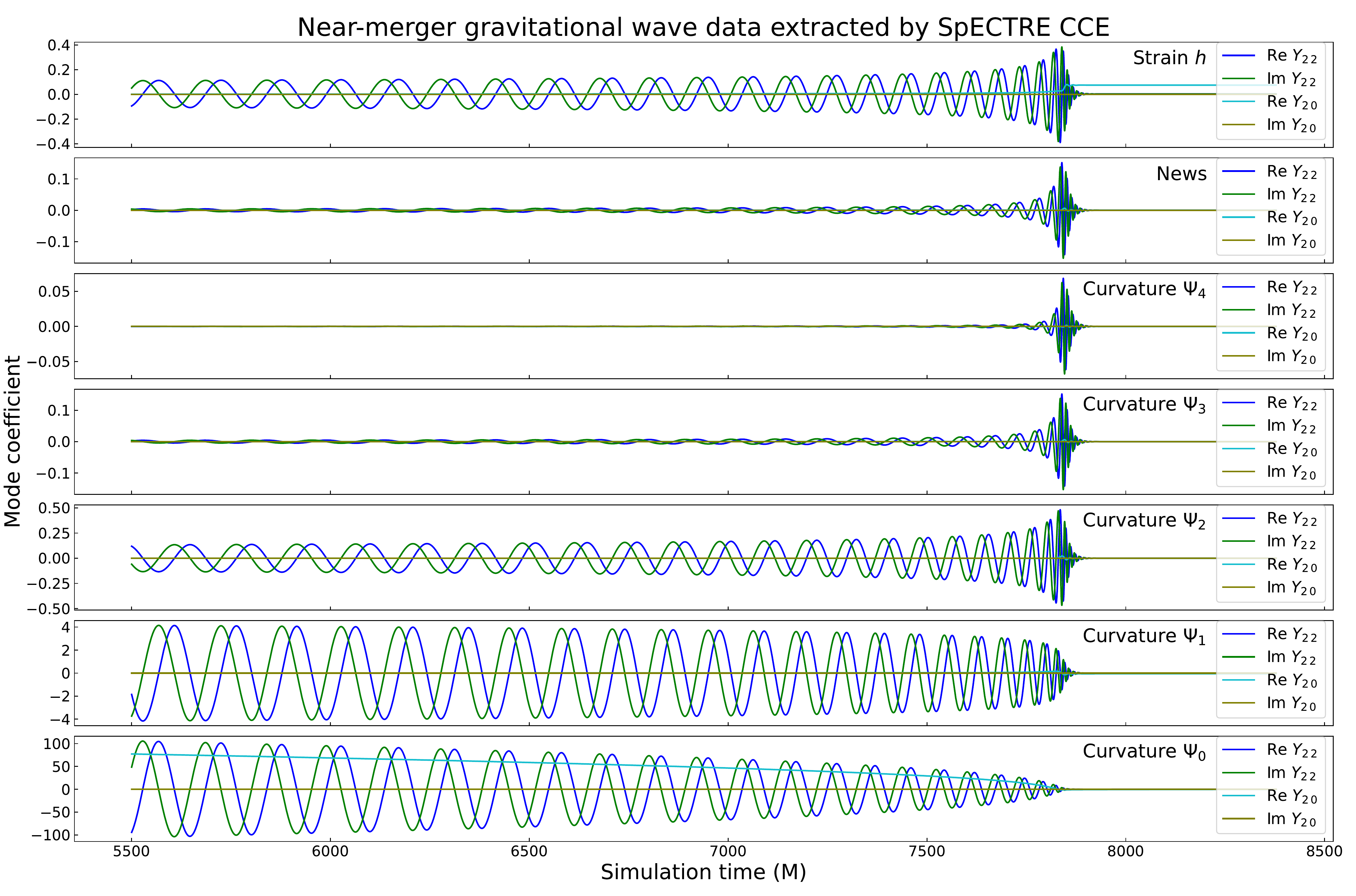}
  \caption{The last several orbits of a waveform extracted using SpECTRE CCE,
    generated from the $R=436$ worldtube of simulation \texttt{SXS:BBH:2096}.}\label{fig:bbh_tests}
\end{figure*}

\begin{figure}
  \includegraphics[width=.48\textwidth]{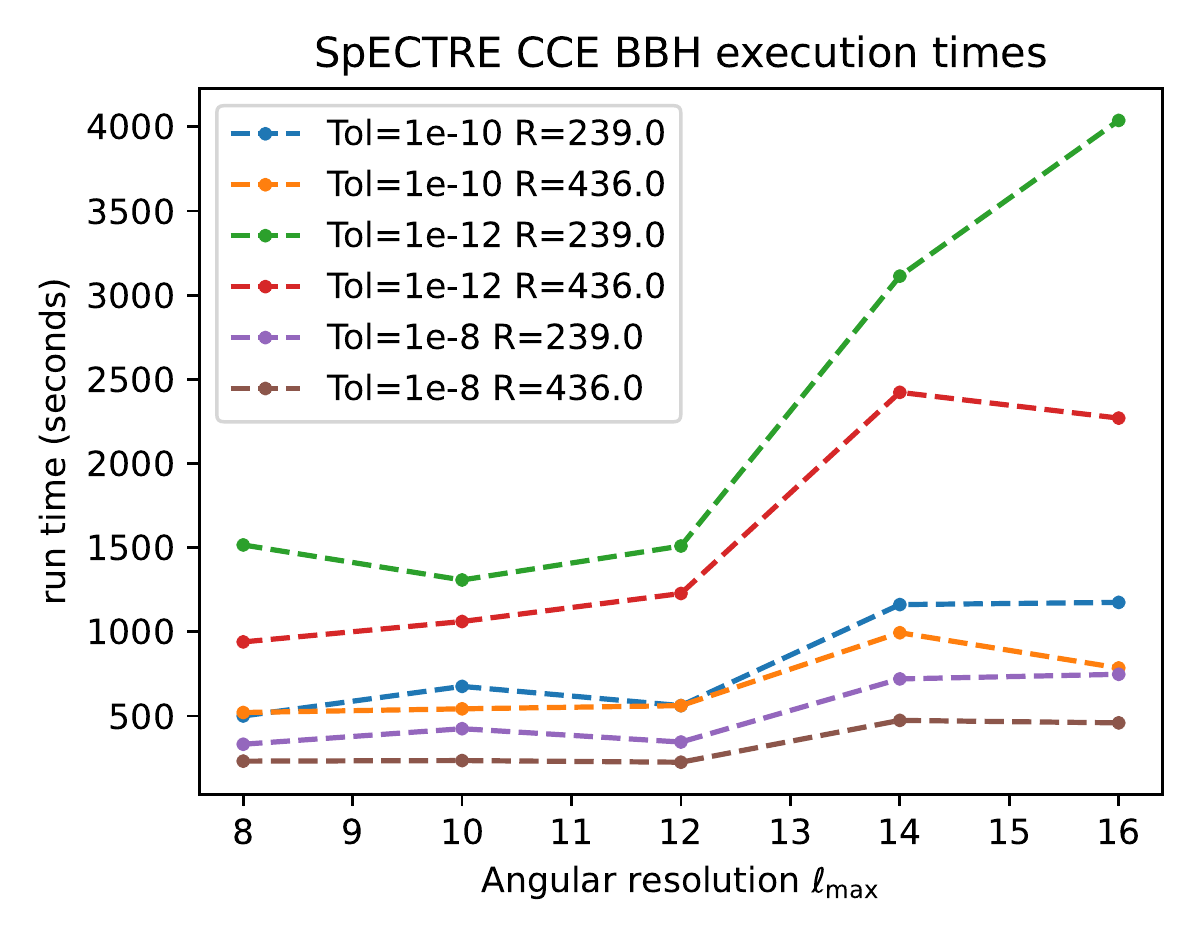}
  \caption{The runtime of SpECTRE CCE applied to the extraction of binary black
    hole worldtube data generated by SpEC for various stepper tolerance targets and
    extraction radii. In practical cases, SpECTRE CCE is able to achieve a high-precision
    wave extraction within 20-40 minutes of runtime.} \label{fig:bbh_performance}
\end{figure}

\begin{figure}
  \includegraphics[width=.48\textwidth]{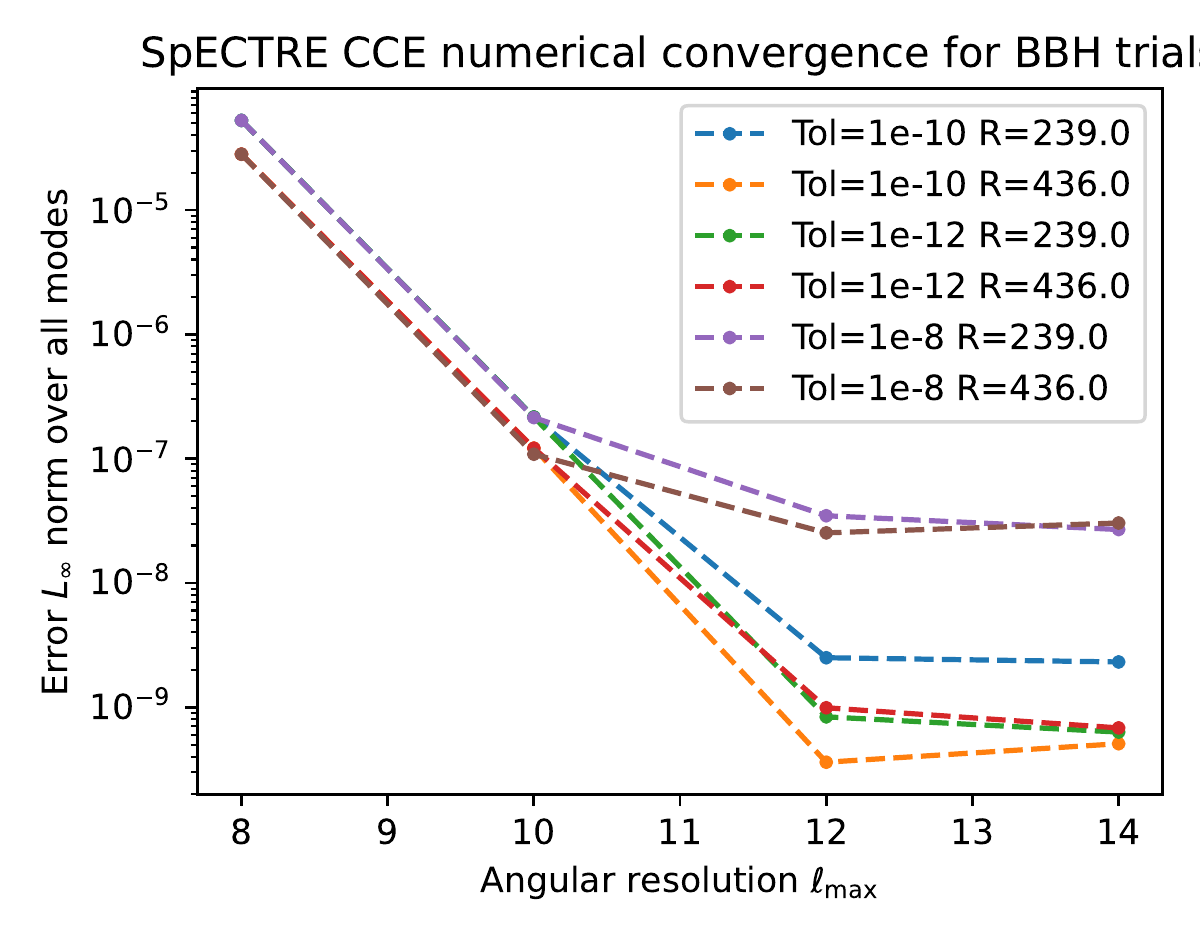}
  \caption{Convergence of the binary black hole trial execution of SpECTRE CCE,
    computed by comparing the extracted news to the value generated at the highest
    angular resolution run, $\ell_{\text{max}} = 16$. The SpECTRE CCE system converges
    rapidly for practical use-cases.} \label{fig:bbh_convergence}
\end{figure}

As the capstone demonstration of the efficacy of the SpECTRE CCE system, we have
performed the full wave extraction of a representative binary black hole
simulation from SpEC.
We have chosen the simulation \texttt{SXS:BBH:2096} from the SXS catalog \cite{Boyle:2019kee, Chu:2015kft}
, for which SXS has stored worldtube data at extraction radii $R=(239, 436, 633, 830)$.

In Fig.~\ref{fig:bbh_tests}, we show the extracted $(2,2)$ and $(2,0)$ modes for each of
the waveform quantities. In Fig.~\ref{fig:bbh_convergence} we show the convergence in numerical
resolution for each of the extraction radii, and in Fig.~\ref{fig:bbh_performance} we show the
performance of the SpECTRE CCE execution for the BBH extractions.
We find that SpECTRE CCE recovers the waveform and Weyl scalars to good
precision, and is able to perform the wave extraction very rapidly, achieving
$\sim 10^{-9}$ residuals with 20-40 minutes of runtime.

\section{Upcoming improvements} \label{sec:upcoming_improvements}

\subsection{Physically motivated initial data generation} \label{sec:improving_initial_data}

The main remaining deficit in the accuracy of the waveforms determined by
SpECTRE CCE is the initial-data transient near the beginning of every
characteristic evolution. 
Without a better method to fix the CCE data on the first hypersurface, waveform
data analysis methods are typically forced to discard the first $\approx3-5$
orbits of the resulting strain waveform, and to correct for the long-lived BMS
frame shift following the initial data transient \cite{Mitman:2021xkq}.
The BMS shift is primarily noticeable in the strain waveform, which displays a
visually apparent offset during the inspiral. Note that
there are also transients (commonly called ``junk radiation'') in the
Cauchy evolution; those transients also force data analysis methods to
discard the beginning of the waveform, but they are not as long-lived as
the CCE transients and dissappear after an orbit or so.

Future work will focus on methods to generate physically motivated data for the
spin-weighted scalar $\breve J$ and angular coordinates $x^A(\hat x^{\hat A})$
on the first CCE hypersurface.       
We anticipate that an improved initial data scheme will construct the state of
the initial hypersurface as an approximation to the system in which the inspiral
had proceeded arbitrarily far into the past of the first Cauchy surface.      
With sufficiently accurate initial data, more of the valuable Cauchy data could
be recovered as high-precision waveform data, and may reduce the demands of the
post-processing BMS gauge transformation.

\subsection{Cauchy-characteristic Matching}

Unlike past implementations of the Cauchy-characteristic evolution system, the
SpECTRE CCE module is able to operate in concert with a simultaneously running
Cauchy simulation with negligible performance degradation.  The key developments
that allow this accomplishment are setting the CCE step size significantly
larger than the Cauchy step as described in Sec.~\ref{sec:independent_step_size}
and numerical improvements described in Sec.~\ref{sec:numerical_optimizations}.

The ability to gracefully run in tandem with a Generalized Harmonic system makes
SPECTRE CCE a well-suited system for an implementation of Cauchy characteristic
matching.  
As derived in \cite{Lindblom:2005qh}, the main piece of information that is
required to determine the ingoing characteristic speeds for the generalized
harmonic system is the Weyl scalar $\Psi_0$, computed in a choice of null
tetrads associated with the Cauchy coordinates.
The value of $\Psi_0$ can be derived directly from spectral operations on any
null hypersurface in the CCE system, and transmitted to the boundary elements in
the GH system to improve the physical boundary condition.

We anticipate that a successful Cauchy-characteristic matching system in SpECTRE
would improve the precision of the outer boundary conditions and reduce
erroneous wave reflections at the outer boundary.  
Further, with sufficient improvement in the boundary behavior, the outer
boundary of the generalized harmonic simulation should be able to be placed at
smaller radii without impacting the waveform precision.
We expect, then, that a smaller simulation domain for the generalized harmonic
system would enable less expensive Cauchy simulations.

\section{Conclusions}

The SpECTRE CCE system represents a significant improvement over previous
methods of performing CCE as well as over
more traditional wave extraction methods.
Our new implementation of CCE is able to rapidly extract waveforms from finished
strong-field simulations or from a simultaneously running Generalized Harmonic
strong-field simulation in SpECTRE. In the latter
case, the SpECTRE CCE system gracefully extracts finalized waveform data in
concert with the strong-field simulation.
Our CCE implementation is extremely fast compared to previous implementations of
CCE \cite{Winicour:1999sg, Bishop:1997ik,Barkett:2019uae}, which provides
significant benefits both for interoperability with other systems and in the
opportunity for users to quickly iterate on new advances in waveform processing
that require the use of CCE as a step in the analysis.

Our implementation takes advantage of recent advances in the understanding of
the formalism underlying the system of Einstein field equations in the
Bondi-Sachs and Bondi-like coordinate systems \cite{Moxon:2020gha}. As a
result, the gauge is specialized to provably avoid any pure-gauge logarithms
that appear in generic Bondi-Sachs-like gauges.
In addition, the implementation is then able to easily compute the
asymptotically leading contribution to all five Weyl scalars (see Fig.~\ref{fig:bbh_tests}).

SpECTRE CCE has already begun to be used to extract valuable insights from
gravitational wave data.
The gains available from highly precise gravitational wave extraction and the
rich data encoded in the Weyl scalars have enabled a number of valuable
early investigations of waveform properties and BMS gauge transformations
\cite{Mitman:2020pbt, Mitman:2021xkq, Foucart:2020xkt, Iozzo:2021vnq}.
We anticipate that precise waveform extraction will play an important role in
the gravitational wave modelling pipeline as next-generation ground-based and
the near future space-based gravitational wave detectors will demand extremely
high quality gravitational wave models \cite{Purrer:2019jcp}.

\section*{Acknowledgments}
We thank Kevin Barkett, Keefe Mitman, and Sizheng Ma for valuable discussions
and suggestions regarding this project.
This work was supported in part by the Sherman Fairchild Foundation
and by NSF Grants No. PHY-1708212, No. PHY-1708213, and No. OAC-1931266
at Caltech and
NSF Grants No. PHY-1912081 and No. OAC-1931280 at Cornell.

\appendix

\section{World tube data representations}

The worldtube metric quantities that the interior Cauchy code must supply to
CCE are all components of the
spatial metric $g_{i j}$, the shift $\beta^i$, the lapse $\alpha$, their radial
derivatives $\partial_r g_{i j}$, $\partial_r \beta^i$, $\partial_r \alpha$, and
their time derivatives $\partial_t g_{i j}$, $\partial_t \beta^i$, and
$\partial_t \alpha$.
This results in a total of
30 tensor components  to store and retrieve.

However, if the initial transformation to Bondi-Sachs coordinates can be
performed before storage, we need only store the boundary values of $\beta, Q,
U, W, H, J, \partial_r J, R$, and $\partial_u R$.
The Bondi-Sachs representation totals 14 real components.
Combined with a representation in spin-weighted spherical harmonics that make
good use of the relationships between $+m$ and $-m$ modes for real functions,
storing Bondi-Sachs data can be a factor of 2--4 cheaper than storing
the full set of metric components and their derivatives.

Because this savings is so great for large catalogs of binary black hole
simulations, SpECTRE also provides a lightweight executable
(\texttt{ReduceCceWorldtube}) for converting inefficient metric component
data to the far smaller Bondi-Sachs data representation.

\section{Interpolation at $\mathcal I^+$} \label{app:scri_interpolation}

The core evolution system using the hierarchical system of CCE equations
\eqref{eq:hierarchy} gives the spin-weighted scalars $\{\breve J, \breve \beta,
\breve U, \breve W, \breve H\}$ that determine the asymptotic partially flat
Bondi-like metric \eqref{eq:bs_metric}.
However, to determine the waveform quantities in asymptotically inertial
coordinates, we must perform a final gauge transformation at $\mathcal I^+$.
Because the partially flat coordinates of the evolution system ensure that the
angular coordinates are asymptotically inertial---their defining equation
\eqref{eq:du_angular_coords} is identical to the asymptotically inertial angular
coordinates constructed in \cite{Bishop:1997ik}---we just need to perform a
coordinate transformation to asymptotically inertial retarded time $\mathring
u(\breve u, \breve x^{\breve A})$.

The asymptotically inertial retarded time $\mathring u$ is determined by
evolving
\begin{equation}\label{eq:du_inertial_retarded_time}
  \partial_{\breve u} \mathring{u} = e^{2 \breve \beta}.
\end{equation}
All other coordinates are identical to their partially flat counterparts
$\mathring r = \hat r$, $\mathring \theta = \hat \theta$, $\mathring \phi = \hat
\phi$.

Once the asymptotically inertial retarded time $\mathring u$ is determined, we
must perform the explicit computations of the asymptotic quantities
\begin{subequations}
  \begin{align}
    h =& \breve{\bar J}^{(1)} + \breve{\bar \eth}^2 \mathring u\\
    N =& e^{-2 \breve \beta^{(0)}}\left[ \breve{\bar{H}}^{(1)} + \breve{\bar \eth}^2 e^{2 \breve \beta^{(0)}}\right]\\
    \Psi_4^{\text{Bondi} (1)} =& \Psi_4^{\text{PF} (1)}\\
    \Psi_3^{\text{Bondi} (2)} =& \Psi_3^{\text{PF} (2)} + \frac{1}{2} \breve \eth \mathring u \Psi_4^{\text{PF} (1)}\\
    \Psi_2^{\text{Bondi} (3)} =& \Psi_2^{\text{PF} (3)} + \breve \eth \mathring u \Psi_3^{\text{PF} (2)} + \frac{1}{4} (\breve \eth \mathring u)^2 \Psi_4^{\text{PF} (1)}\\
    \Psi_1^{\text{Bondi} (4)} =& \Psi_1^{\text{PF} (4)} + \frac{3}{2} \breve \eth \mathring u \Psi_2^{\text{PF}(3)} + \frac{3}{4} (\breve \eth \mathring u)^2 \Psi_3^{\text{PF}(2)} \notag\\&+ \frac{1}{8} (\breve \eth \mathring u)^3 \Psi_4^{\text{PF}(1)}\\
    \Psi_0^{\text{Bondi} (5)} =& \Psi_0^{\text{PF} (5)} + 2 \breve \eth \mathring u \Psi_1^{\text{PF} (4)} + \frac{3}{4}(\breve \eth \mathring u)^2 \Psi_2^{\text{PF} (3)} \notag\\
       &+ \frac{1}{2}(\breve \eth \mathring u)^3 \Psi_3^{\text{PF}(2)} + \frac{1}{16} (\breve \eth \mathring u)^4 \Psi_4^{\text{PF}(1)}.
  \end{align}
\end{subequations}

\begin{figure}
  \includegraphics[width=.36\textwidth]{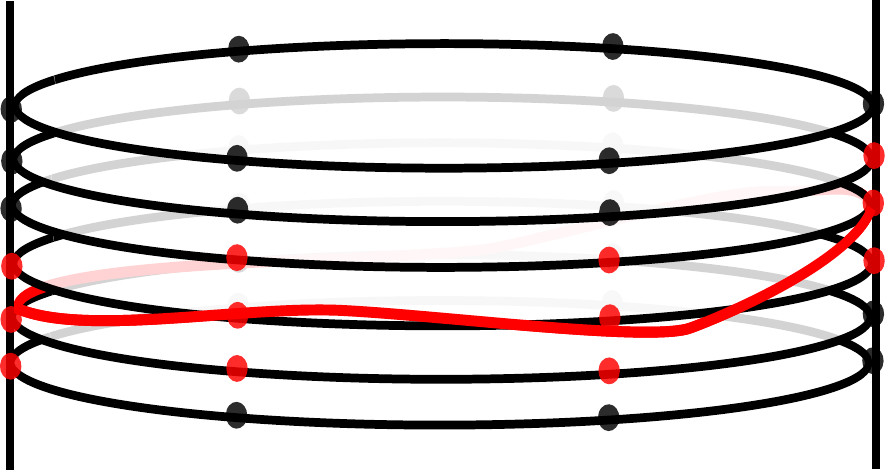}
  \caption{A sketch of the interpolation required at $\mathcal I^+$. The black
    rings 
    represent the time series of spherical surface data produced by
CCE at $I^+$ and the points represent the collocation points on which the field
values are provided. 
The red curve represents a single value of asymptotically inertial time
$\mathring u(\breve u, \breve x^{\breve A})$ on which we wish to evaluate the
waveform. The red points  are those
we would use to perform a
second-order interpolation to the asymptotically inertial time. Note that we may
need to use different sets of source time values $\breve u$ at different
collocation points.}\label{fig:scri_interpolation_sketch}
\end{figure}

However, once we've computed these waveform quantities, we still need to perform
the interpolation to cuts of $\mathcal I^+$ at constant inertial retarded time
$\mathring u$.
To perform the interpolation, we record several time steps of the CCE evolution,
until we have sufficient data at $\mathcal I^+$ to perform a barycentric
rational \cite{Floater:2007} 
interpolation to the target cut of constant $\mathring u$.
This process is illustrated in Fig.~\ref{fig:scri_interpolation_sketch}.

\section{Clenshaw recurrence details} \label{sec:clenshaw_details}

\subsubsection{Spin-weighted spherical harmonics in terms of Jacobi polynomials}

A number of representation choices exist for the spin-weighted spherical
harmonic basis, most of which are related by phase and sign changes.
For simplicity, we take the definition of the spin-weighted spherical
harmonics ${}_s Y_{\ell m}(\theta, \phi)$ directly in terms of the Wigner rotation
matrices $d^l_{-m, s}(\theta)$ \cite{Goldberg1966uu},
\begin{equation} \label{eq:SwshToWigner}
  {}_s Y_{\ell m}(\theta, \phi) = (-1)^m \sqrt{\frac{2 \ell + 1}{4 \pi}}
  e^{i m \phi} d^{\ell}_{-m, s}(\theta)
\end{equation}

The Wigner rotation matrices $d^{\ell}_{-m, s}$ may then be expressed in terms of
Jacobi polynomials \cite{Varshalovich:1988}. Define:
\begin{subequations}
\begin{align}
  a &= |s + m| \\
  b &= |s - m| \\
  k &= -\frac{1}{2}(a + b)\\
  \lambda &= \begin{cases}0, & s \ge -m,\\
    s + m, & s < -m. \end{cases}
\end{align}
\end{subequations}
Then,
\begin{align} \label{eq:WignerToJacobi}
  d^{\ell}_{-m, s}(\theta)
  =&  (-1)^\lambda \sqrt{\frac{(\ell + k)! (\ell + k + a + b)!}
    {(\ell + k + a)!(\ell + k + b)!}}\notag \\ &{}\times\sin^a\left( \frac{\theta}{2}\right)
  \cos^b\left(\frac{\theta}{2}\right) P^{(a, b)}_{\ell + k}(\cos\theta).
\end{align}
We have deviated from the notation of \cite{Varshalovich:1988} and others by
separating out the $\ell$ contribution from the variable $k$.
With the notation in (\ref{eq:WignerToJacobi}), all dependence on $\ell$ is
explicit. This assists the derivation below  of recurrence relations
for successive $\ell$ at fixed $s$ and $m$ (and so also fixed $a$, $b$, and
$k$).

Note that for any particular  $(s,m)$, the lowest $\ell$-mode that is required is
$\ell_{\min} = \max(|m|, |s|)$, and that $k = \min(-|m|, -|s|)$.
Thus each of the
recurrence relations has its
lowest two contributions determined by $P_0^{(a, b)}$
and $P_1^{(a,b)}$, which have convenient closed forms:
\begin{subequations}
  \begin{align}
    P_0^{(a, b)}(\cos\theta) &= 1\\
    P_1^{(a, b)}(\cos\theta) &= (a + 1) + (a + b + 2) \frac{\cos\theta - 1}{2}
  \end{align}
\end{subequations}

For reference, the three-term recursion relation we use for the Jacobi polynomials is
\cite{Bateman:1953},
\begin{widetext}
\begin{subequations} \label{eq:JacobiRecurrance}
\begin{align}
  P_n^{(a, b)}(x)
  &= {}_P \alpha_n^{(a, b)}(x) P_{n-1}(x) + {}_P \beta_n^{(a, b)} P_{n-2}(x)\\
  {}_P \alpha_n^{(a,b)}(x)
  &= \frac{2n + b + a - 1}{2n (n + a + b)}\left[(2n + a + b) x
    + \frac{(a^2 - b^2)}{2n + a + b - 2}\right]\\
  {}_P \beta_n^{(a,b)}
  &= \frac{-(2n + a + b)(n + a - 1)(n + b - 1)}{n (n + a + b) (2n + a + b - 2)}.
\end{align}
\end{subequations}
\end{widetext} In (\ref{eq:JacobiRecurrance}), we denote the recurrence
coefficients with a leading subscript $P$, to avoid ambiguity with other
recurrence coefficients in this paper.

\subsubsection{Recursion relations for application of Clenshaw algorithm}

In general, to perform a spin-weighted spherical harmonic interpolation from a
prescribed set of collocation points, one first performs a transformation
to
spin-weighted coefficients ${}_s a_{\ell m}$, then interpolates to each desired
$(\theta_i, \phi_i)$ by evaluating the sum
\begin{equation}
  f(\theta_i, \phi_i) = \sum_{m = -\ell_{\max}}^{\ell_{\max}}
  \sum_{\ell=\min(|m|, |s|)}^{\ell_{\max}} {}_s a_{\ell m}\, {}_s Y_{\ell m}(\theta_i, \phi_i).
\end{equation}
The Clenshaw-based algorithm will possess an outer loop over
$m \in [-\ell_{\max}, \ell_{\max}]$ modes for a given spin $s$.
The inner sum will then be evaluated using the Clenshaw recurrence algorithm
described in section \ref{sec:clenshaw}.
For this section, we focus on the formulas necessary to apply the Clenshaw
algorithm to the innermost loop.

For spin-weighted spherical harmonics with $\ell \ge \max(|m|, |s|) + 2$,
we seek a recurrence relation of the form
\begin{align} \label{eq:swsh_recurrence}
  {}_s Y_{\ell m}(\theta, \phi)
  =& {}_Y \alpha_{\ell}^{(a,b)}(\theta, \phi) {}_s Y_{\ell-1\, m}(\theta, \phi)\notag\\
  &+ {}_Y \beta_{\ell}^{(a, b)} {}_s Y_{\ell - 2\, m}(\theta, \phi).
\end{align}
The coefficients in \eqref{eq:swsh_recurrence}  are labeled with
  a leading subscript $Y$, and may be inferred from the relation
between the spin-weighted spherical harmonics and the Jacobi polynomials.
The result is the recurrence coefficients
\begin{widetext}
\begin{subequations}
\begin{align}
  {}_Y \alpha_{\ell}^{(a, b)}
  =& \sqrt{\frac{2 \ell + 1}{2 \ell - 1}}\sqrt{\frac{(\ell + k)(\ell + k + a + b)}{(\ell + k + a)(\ell + k + b)}}
    \, {}_P \alpha_{\ell + k}^{(a, b)}(\cos\theta)\notag \\
  =& \sqrt{\frac{2 \ell + 1}{2 \ell - 1}} \frac{2(\ell + k) + b + a - 1}
    {2 \sqrt{(\ell + k)(\ell + k + a + b)(\ell + k + a)(\ell + k + b)}}\notag\\
    &\times \left[(2(\ell+k) + a + b) \cos\theta
    + \frac{a^2 - b^2}{2(\ell+k) + a + b - 2}\right],\\
  {}_Y \beta_{\ell}^{(a, b)}
  =& \sqrt{\frac{2 \ell + 1}{2 \ell - 3}}\sqrt{\frac{(\ell + k)(\ell + k - 1)(\ell + k + a + b)(\ell + k + a + b - 1)}
    {(\ell + k + a)(\ell + k + a - 1)(\ell + k + b)(\ell + k + b - 1)}}
    \, {}_P \beta_{\ell + k}^{(a, b)} \notag\\
  =& -\sqrt{ \frac{(2 \ell + 1)(\ell + k + a - 1)(\ell + k + b - 1)(\ell + k - 1)(\ell + k + a + b - 1)}
    {(2 \ell - 3)(\ell + k)(\ell + k + a + b)(\ell + k + a)(\ell + k + b)}}\frac{2(\ell + k) + a + b}{2(\ell + k) + a + b - 2}.
\end{align}
\end{subequations}
\end{widetext} When generating the ${}_Y \alpha$ and ${}_Y \beta$ coefficients,
it would be wasteful to populate a full two-dimensional space of integers for
$(a, b)$ values.
For each spin $s$, there is a one-to-one mapping between $m$ and $(a, b)$ pairs,
so for each spin value, exactly $2 \ell_{\max} + 1$ recurrence coefficient sets
should be generated.
For each coefficient set, coefficients are needed with indices
$ \ell \in [\min(|m|, |s|), \ell_{\max}]$.

\subsubsection{Relations between successive iterations for spin-weighted spherical harmonics}

In this section, we describe the computations necessary to obtain the two lowest
spin-weighted spherical harmonics for each Clenshaw recurrence evaluation, and
suggest a method by which several of these explicit functions may also be
determined by recurrence in $m$ to limit evaluations of the factorial prefactor
and powers of trigonometric functions found in (\ref{eq:WignerToJacobi}).

First, we note that it is desirable to first evaluate the recurrence for $m=0$,
then perform the sequence of positive $m$ and negative $m$ as further branches.
The reason for this evaluation structure is based on the observation that
successive factors of $\sin^a(\theta/2) \cos^b(\theta/2)$ can be
computed from previous ones provided that $a$ and $b$ both increment from one
step to the next.
From (\ref{eq:abValues}), we see that that this will be true for ascending
values of $|m|$ whenever $|s| < |m|$.

Therefore, it is possible to recursively obtain ${}_s Y_{|m|\, m}$ and ${}_s
Y_{|m| + 1\, m}$ from the previously determined harmonics, for all harmonics
where $|m| > |s|$.
Recursive construction won't be available for complete generality, as the steps
for which $|s| \ge |m|$ involve both the increase and decrease of $a$'s and
$b$'s, so would involve the potentially dangerous division by
$\sin(\theta/2)$, which is ill-defined near one pole.
For those $2 |s|$ steps, our algorithm accepts the cost of the direct evaluation
of the (small) powers.
For most realistic applications, the value $|s|$ will be far smaller than
$\ell_{\max}$, so most $m$ steps 
can be evaluated by the recursive method.

In particular, for $|m| > |s|$, the first required explicit function for the
Clenshaw recurrence is
\begin{widetext}
\begin{equation} \label{eq:MRecursion}
  {}_s Y_{|m|\, m} = (-1)^{\Delta \lambda} \sqrt{\frac{2 |m| + 1}{2 |m| - 1}
    \frac{(\ell + k + a + b  - 1)(\ell + k + a + b)}
    {(\ell + k + a)(\ell + k + b)}} e^{i \phi}
  \sin\left(\theta / 2\right) \cos\left(\theta / 2 \right)
  \begin{cases}
    {}_s Y_{|m| - 1\, m - 1}, & m > 0,\\
    {}_s Y_{|m| - 1\, m + 1}, & m < 0,
  \end{cases}
\end{equation}
\end{widetext}
where the difference $\Delta \lambda = \lambda_m - \lambda_{\pm m}$ is
sufficiently cheap to compute on a case-by-case basis.
Finally, the second harmonic can always be easily evaluated from the first
harmonic of a sequence.
The simple relation arises from noting that the first Jacobi polynomial in each
sequence is unity.
Therefore, computation can once again be saved in determining
${}_s Y_{\ell_{\min} + 1, m}$:
\begin{widetext}
\begin{equation} \label{eq:SecondL}
  {}_s Y_{\ell_{\min} + 1\, m} = \sqrt{\frac{2 \ell_{\min} + 3}{ 2 \ell_{\min} + 1}}
  \sqrt{\frac{(\ell_{\min} + k + 1)(\ell_{\min} + k + a + b + 1)}
    {(\ell_{\min} + k + a + 1)(\ell_{\min} + k + b + 1)}}
  \left[(a + 1) + (a + b + 2) \frac{(\cos\theta - 1)}{2}\right]
  {}_s Y_{\ell_{\min}\, m}.
\end{equation}
\end{widetext}

With the above recurrence for the successive starting $m$ modes, it is only
necessary to evaluate ${}_s Y_{\ell_{\min}\, m}$ for $|m| \le |s|$ from the explicit
formulas for the Wigner rotation matrices (\ref{eq:WignerToJacobi}).

\bibliography{cce_numerics}

\bibliographystyle{apsrev4-2}

\end{document}